\documentclass[11pt,a4paper]{article}
\usepackage{jcappub}

\title{Observational constraints on scalar field models of dark energy with barotropic equation of state}

\author[a]{Olga Sergijenko,}
\author[b]{Ruth Durrer}
\author[a]{and Bohdan Novosyadlyj}
\affiliation[a]{Astronomical Observatory of Ivan Franko National University of Lviv, Kyryla i Methodia str. 8, Lviv, 79005, Ukraine}
\affiliation[b]{Universit\'e de Gen\`eve, D\'epartement de Physique Th\'eorique and CAP, 24 quai Ernest-Ansermet, CH-1211 Gen\`eve 4, Switzerland}
\emailAdd{olka@astro.franko.lviv.ua}
\emailAdd{ruth.durrer@unige.ch}
\emailAdd{novos@astro.franko.lviv.ua}

\abstract{We constrain the parameters of dynamical dark energy in the form of a classical or tachyonic scalar field with barotropic equation of state jointly with other cosmological parameters using the following combined datasets: the CMB power spectra from WMAP7, the baryon acoustic oscillations in the space distribution of galaxies from SDSS DR7, the power spectrum of luminous red galaxies from SDSS DR7 and the light curves of SN Ia from 2 different compilations: Union2 (SALT2 light curve fitting) and SDSS (SALT2 and MLCS2k2 light curve fittings). It has been found that the initial value of dark energy equation of state parameter is constrained very weakly by most of the data while the other cosmological parameters are well constrained: their likelihoods and posteriors are similar, their forms are close to Gaussian (or half-Gaussian) and their confidence ranges are narrow. The most reliable determinations of the best-fit value and $1\sigma$ confidence range for the initial value of the dark energy equation of state parameter are obtained from the combined datasets including SN Ia data from the full SDSS compilation with MLCS2k2 fit light curves. In all  cases the best-fit value of this parameter is lower than the value of corresponding parameter for current epoch. Such dark energy loses its repulsive properties and in future the expansion of the Universe changes into contraction.
  
We also perform a forecast for the Planck mock data and show that they narrow significantly the confidence ranges of cosmological parameters values, moreover, their  combination with SN SDSS compilation with MLCS2k2 light curve fitting may exclude the fields with initial equation of state parameter $>-0.1$ at 2$\sigma$ confidence level.}


\begin{document}

\maketitle

\section{Introduction}

The discovery of the accelerated expansion of the Universe has led to the introduction of a new mysterious component -- dark energy. Its unknown nature is one of the main puzzles of modern cosmology. Since the simplest explanation -- a cosmological constant, faces many interpretational problems, a variety of alternative models have been proposed (see reviews \cite{rev1}-\cite{rev8} and books \cite{tb1}-\cite{tb3}). The simplest alternative approach treats dark energy as a scalar field with standard Lagrangian. The model  is defined by the scalar field potential which can be either physically motivated or obtained via reverse engineering from the variables describing dark energy in a phenomenological fluid approach: the energy density and equation of state (EoS) parameter \textit{w}. The latter can either be constant or vary in time. The character of the temporal variation of \textit{w} is usually assumed ad hoc. Nevertheless, physically motivated dependences of the equation of state on time are sought.

The simplest and most widely used Lagrangians of a scalar field are the Klein-Gordon (called also classical) and the Dirac-Born-Infeld (often called tachyon) ones.

In our previous papers \cite{Novosyadlyj2010a}-\cite{Novosyadlyj2010c} we have analyzed the parametrization of the EoS by its current $w_0$ and constant adiabatic sound speed $c_a^2$, which corresponds to the EoS parameter at the beginning of expansion $w_e$. In this case the dark energy EoS is of the generalized linear barotropic form. Such a parametrization is easy to  motivate physically. For the dark energy in the form of the scalar fields with barotropic EoS the analytical solutions for the field variables and potentials exist for both classical and tachyonic Lagrangians. In the case of these dark energy models the relation between the current and early values of EoS parameter determines two drastically different scenarios for the future evolution of the Universe. 

The only way to verify the plausibility of a dark energy model is to confront its predictions with the observational data and to find the allowed ranges of its parameters (for discussion of the cosmological parameter estimation see e. g. \cite{tb1}). For this purpose the Markov Chain Monte-Carlo approach is widely used. In the paper \cite{Novosyadlyj2010a} we have found that the early EoS parameter $w_e$ remains unconstrained by two combined datasets including the recent data on CMB anisotropy, large-scale structure of the Universe and light curves of supernovae type Ia. This is due to the significant non-Gaussianity of the likelihood function with respect to $w_e$. In order to find the best-fit value of this parameter along with the best-fit values of the remaining cosmological parameters other datasets should be extensively analyzed. 

The goal of this paper is to derive constraints on the parameters of models with scalar fields using different current and near future data and to present observational constraints on cosmological models with classical and tachyonic scalar fields with barotropic equation of state as dark energy.

The paper is organized as follows. In Section \ref{theory} we discuss the parametrization of barotropic EoS parameter by its current and early values, the evolution of the scale factor in single- and multicomponent models and the potentials of classical and tachyonic scalar fields with barotropic EoS. In Section \ref{curoc} we present the observational constraints on the parameters defining the barotropic EoS obtained from the currently available data. In Section \ref{futoc} we forecast the precision, with which the expected Planck data will be able to constrain the cosmological parameters of models with scalar field dark energy with barotropic EoS. The conclusion are given in Section \ref{concl}.  

\section{Scalar field models of dark energy with barotropic equation of state}\label{theory}

The background Universe is assumed to be spatially flat, homogeneous and isotropic with Friedmann-Robertson-Walker (FRW) metric of 4-space $$ds^2=g_{ij} dx^i dx^j =a^2(\eta)(d\eta^2-\delta_{\alpha\beta} dx^{\alpha}dx^{\beta}),$$ where $\eta$ is the conformal time defined by $dt=a(\eta)d\eta$ and $a(\eta)$ is the scale factor, normalized to 1 at the current epoch (here and below we put $c=1$). The Latin indices \textit{i, j,...} run from 0 to 3 and the Greek ones are used for the spatial part of the metric: $\alpha,\,\beta,..=1,2,3$.

We consider a multicomponent model of the Universe filled with non-relativistic particles (cold dark matter and baryons), relativistic particles (thermal electromagnetic radiation and massless neutrino) and minimally coupled dark energy. The dark energy is assumed to be a scalar field with either Klein-Gordon (classical, below: CSF) or Dirac-Born-Infeld (tachyonic, below: TSF) Lagrangian 
\begin{eqnarray}
L_{clas}=X-U(\phi),\,\,\,
L_{tach}=-\tilde{U}(\xi)\sqrt{1-2\tilde{X}},\label{L}
\end{eqnarray}
where ${U}(\phi)$ and $\tilde{U}(\xi)$ are the field potentials defining the model of the scalar field, $X=\phi_{;i}\phi^{;i}/2$ and $\tilde{X}=\xi_{;i}\xi^{;i}/2$ are kinetic terms. We assume the homogeneity of background scalar fields ($\phi(\textbf{x},\eta)=\phi(\eta)$, $\xi(\textbf{x},\eta)=\xi(\eta)$), so that their energy density and pressure depend only on time:
\begin{eqnarray}
&&\rho_{clas}=X+{U}(\phi),\,\,\,\,\,
p_{clas}=X-{U}(\phi),\\
&&\rho_{tach}=\frac{\tilde{U}(\xi)}{\sqrt{1-2\tilde{X}}},\,\,\,\,
p_{tach}=-\tilde{U}(\xi)\sqrt{1-2\tilde{X}}.
\end{eqnarray}
The EoS parameters $w_{de}\equiv p_{de}/\rho_{de}$ for these fields are
\begin{eqnarray}
 w_{clas}=\frac{X-U}{X+U},\,\,\,\,\,w_{tach}=2\tilde{X}-1\label{w_X}.
\end{eqnarray}
Using the last relations the field variables and potentials can be presented in terms of densities and EoS parameters as:
\begin{eqnarray}
&&\phi(a)-\phi_0=\pm\int_1^a\frac{da'\sqrt{\rho_{de}(a')(1+w(a'))}}{a'H(a')},\label{vcl}\\
&&U(a)=\frac{\rho_{de}(a)\left[1-w(a)\right]}{2}\label{pcl}
\end{eqnarray}
for the classical Lagrangian and
\begin{eqnarray}
&&\xi(a)-\xi_0=\pm\int_1^a\frac{da'\sqrt{1+w(a')}}{a'H(a')},\label{vt}\\
&&\tilde{U}(a)=\rho_{de}(a)\sqrt{-w(a)}\label{pt}
\end{eqnarray}
for the tachyonic case.

The dynamics of expansion of the Universe is fully described by the Einstein equations
\begin{eqnarray}
R_{ij}-{\frac{1}{2}}g_{ij}R=8\pi G \left(T_{ij}^{(m)}+T_{ij}^{(r)}+T_{ij}^{(de)}\right),
\label{E_eq}
\end{eqnarray}
where $R_{ij}$ is the Ricci tensor and $T_{ij}^{(m)},\,T_{ij}^{(r)},\,T_{ij}^{(de)}$ are the energy-momentum tensors of non-relativistic matter (m), relativistic matter (r), and dark energy (de) correspondingly. Assuming that the interaction between these components is only gravitational, each of them should satisfy the differential energy-momentum conservation law separately, which for a perfect fluid with density $\rho_n$ and pressure $p_n$ related by the equation of state $p_n=w_n\rho_n$ yields:
\begin{eqnarray}
a\rho'_{n}=-3\rho_{n}(1+w_{n})\label{rho'},
\end{eqnarray}
here and below a prime denotes the derivative with respect to the scale factor $a$. For the non-relativistic matter $w_{m}=0$ and $\rho_{m}=\rho_{m}^{(0)}a^{-3}$, for the relativistic one $w_{r}=1/3$ and $\rho_{r}=\rho_{r}^{(0)}a^{-4}$. Hereafter ``0'' denotes the current values.

The EoS parameter $w$ and the adiabatic sound speed $c_a^2\equiv\dot{p}_{de}/\dot{\rho}_{de}$ of dark energy are related by the differential equation:
\begin{eqnarray}
 aw'=3(1+w)(w-c_a^2),\label{w'}
\end{eqnarray}
Note that in the case of studied types of dark energy the adiabatic sound speed is not the true velocity of sound propagation. The scalar fields have non-negligible entropy and thus non-adiabatic pressure perturbations. In the dark energy rest frame the total pressure perturbation can be presented as $\delta p_{de}=c_s^2\delta_{de}$, where the effective sound speed $c_s^2\equiv p,_{X}/\rho,_{X}=L,_{X}/(2XL,_{XX}+L,_{X})$ is different for the classical and tachyonic scalar fields (1 and $-w$ correspondingly). Therefore, the adiabatic sound speed is a useful quantity having the form of corresponding thermodynamical quantity. Below we will see its real physical meaning.
In general $c_a^2$ can be a function of time, but here we assume it to be constant: $c_a^2=const$. In this case the time derivative of $p_{de}$ is proportional to the time derivative of $\rho_{de}$ or in integral form:
\begin{eqnarray}
 p_{de}=c_a^2\rho_{de}+C,\label{beos}
\end{eqnarray}
where $C$ is a constant. The above expression is the generalized linear barotropic equation of state.
The solution of equation (\ref{w'}) for $c_a^2=const$ is
\begin{eqnarray}
 w(a)=\frac{(1+c_a^2)(1+w_0)}{1+w_0-(w_0-c_a^2)a^{3(1+c_a^2)}}-1,\label{w}
\end{eqnarray}
where the integration constant $w_0$ is the current value of $w$. One can easily find that (\ref{w}) gives (\ref{beos}) with $C=\rho_{de}^{(0)}(w_0-c_a^2)$, where $\rho_{de}^{(0)}$ is the current density of dark energy. Substituting (\ref{w}) into (\ref{w'}) we see that for quintessence fields ($w_0>-1$) the derivative of EoS parameter with respect to the scale factor is negative for $c_a^2>w_0$ and positive for $c_a^2<w_0$.

Therefore, we have two values $w_0$ and $c_a^2$ defining the EoS parameter $w$ for any scale factor $a$. As it follows from (\ref{w}), parameter $c_a^2$ corresponds to the EoS parameter at the beginning of expansion: $w_e\equiv w(0)\equiv c_a^2$. Below we will call this parameter $w_e$ instead of $c_a^2$ as such notation reflects better its physical meaning. The differential equation (\ref{rho'}) with $w$ in the form (\ref{w}) has the analytic solution
\begin{eqnarray}
\rho_{de}=\rho_{de}^{(0)}\left(\frac{(1+w_0)a^{-3(1+w_e)}+w_e-w_0}{1+w_e}\right).\label{rho}
\end{eqnarray}
Using the dependences of densities of each component on the scale factor the following equations for background dynamics can be deduced from the Einstein equations (\ref{E_eq}):
\begin{eqnarray}
 H&=&H_0\sqrt{\Omega_r/a^{4}+\Omega_m/a^{3}+\Omega_{de}f(a)}, \label{H}\\
 q&=&\frac{1}{2}\frac{2\Omega_r/a^{4}+\Omega_m/a^{3}+(1+3w)\Omega_{de}f(a)}
{\Omega_r/a^{4}+\Omega_m/a^{3}+\Omega_{de}f(a)},\label{q}
\end{eqnarray}
where $f(a)=[(1+w_0)a^{-3(1+w_e)}+w_e-w_0]/(1+w_e)$. Here $H\equiv\dot{a}/{a^2}$ is the Hubble parameter (expansion rate) and $q\equiv-\left(a\ddot{a}/\dot{a}^2-1\right)$ is the acceleration parameter (``$\;\dot{}\;$''$\equiv\partial/\partial\eta$). The equations (\ref{H})-(\ref{q}) completely describe the dynamics of expansion of the homogeneous and isotropic Universe.

In our previous paper \cite{Novosyadlyj2010a} we have analyzed in detail three possible scenarios of the future evolution of the Universe. For $w_e>w_0$  the dark energy will tend to mimic a cosmological constant in the future, such a Universe will expand forever as in de-Sitter inflation. For $w_e=w_0$ (the simplest case) the future of the Universe is eternal power-law expansion. For $w_e<w_0$ the dark energy turns away from its repulsive properties and in the future the expansion of the Universe will turn into contraction.

For a realistic multicomponent model it is possible to find the time dependence of the scale factor from (\ref{H}) only numerically. However let us consider the simple scalar field model of spatially-flat Universe filled only with the scalar field with barotropic EoS. In this case the equation has the analytical solutions for the evolution of scale factor
\begin{eqnarray*}
a(t)=\left(\frac{1+w_0}{w_e-w_0}\right)^{\frac{1}{3\left(1+w_e\right)}}\sinh^{\frac{2}{3\left(1+w_e\right)}}\left(\frac{3}{2}\sqrt{\left(1+w_e\right)\left(w_e-w_0\right)}H_0t\right)
\end{eqnarray*}
for $w_e>w_0$ and
\begin{eqnarray*}
a(t)=\left(\frac{1+w_0}{w_0-w_e}\right)^{\frac{1}{3\left(1+w_e\right)}}\sin^{\frac{2}{3\left(1+w_e\right)}}\left(\frac{3}{2}\sqrt{\left(1+w_e\right)\left(w_0-w_e\right)}H_0t\right)
\end{eqnarray*}
for $w_e<w_0$. For $w_e>w_0$ this solution has been found previously in \cite{Gorini2003}, the corresponding expression for $w_e<w_0$ can be easily obtained from it.

\begin{figure}
\centerline{\includegraphics[width=0.65\textwidth]{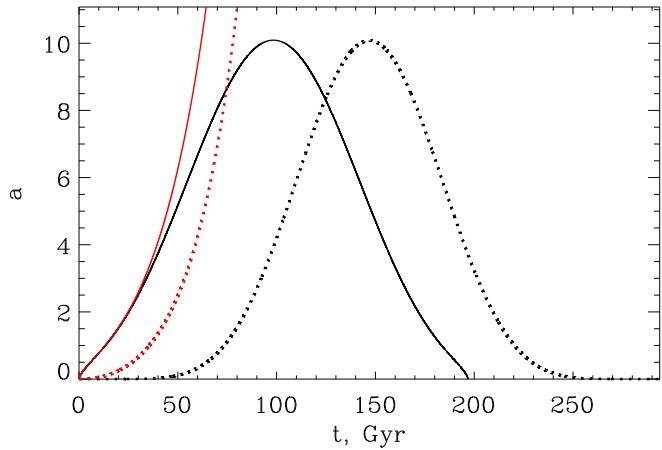}} 
\caption{Evolution of the scale factor in cosmological models with scalar fields with barotropic EoS as dark energy. Black lines: $w_e<w_0$, red: $w_e>w_0$. Solid lines -- numerical solutions for the multicomponent models, dotted -- analytical solutions for the simple scalar field models.}
\label{fig1}
\end{figure}

In Fig.~\ref{fig1} the numerical solutions for the multicomponent model and the corresponding analytical ones for the simple scalar field model are presented for both cases: $w_e>w_0$ and $w_e<w_0$. We see that indeed there exist two possible scenarios for the  future evolution of the Universe. In the case $w_e>w_0$ in far future the Universe will experience eternal asymptotically de-Sitter expansion, while in the case $w_e<w_0$ the cosmological expansion will slow down reaching the turnaround time after which the Universe will collapse. So, in the latter case the whole history of the Universe is limited in time. The difference between the corresponding curves in single-multicomponent models arises from the fact that in the multicomponent Universe at the early stages of expansion the relativistic and non-relativistic matter dominate. At scale factors corresponding to the dark energy domination in the multicomponent model the shapes of both curves become similar but are shifted in time.

\begin{figure}
\centerline{\includegraphics[width=\textwidth]{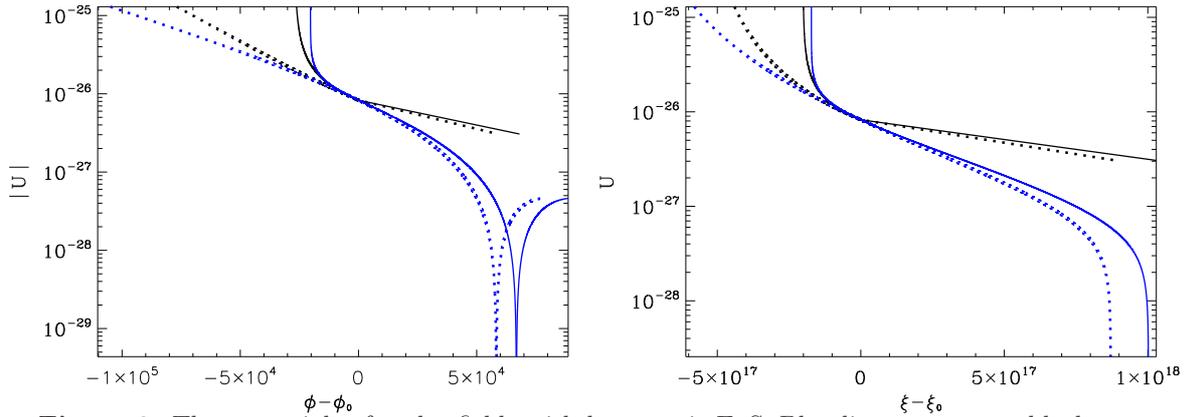}} 
\caption{The potentials of scalar fields with barotropic EoS. Blue lines: $w_e<w_0$, black: $w_e>w_0$. Solid lines -- numerical solutions for multicomponent models, dotted -- analytical solutions for simple scalar field models. Left: CSF, right: TSF.}
\label{fig2}
\end{figure}

In the simple scalar field model from (\ref{vcl})-(\ref{pt}) it is easy to obtain the analytical expressions for the field potentials. For CSF they read: 
\begin{eqnarray*}
&&U(\phi-\phi_0)=\frac{3H_0^2}{8\pi G}\frac{w_e-w_0}{1+w_e}+\frac{3H_0^2}{8\pi G}\frac{1-w_e}{1+w_e}\frac{w_e-w_0}{2}\\
&&\times\sinh^2\left(\sqrt{6\pi G}\sqrt{1+w_e}\left(\phi-\phi_0\right)-\coth^{-1}\left(\sqrt{\frac{1+w_e}{1+w_0}}\right)\right)
\end{eqnarray*}
in the case of $w_e>w_0$ and
\begin{eqnarray*}
&&U(\phi-\phi_0)=\frac{3H_0^2}{8\pi G}\frac{w_e-w_0}{1+w_e}+\frac{3H_0^2}{8\pi G}\frac{1-w_e}{1+w_e}\frac{w_0-w_e}{2}\\
&&\times\cosh^2\left(\sqrt{6\pi G}\sqrt{1+w_e}\left(\phi-\phi_0\right)-\tanh^{-1}\left(\sqrt{\frac{1+w_e}{1+w_0}}\right)\right)
\end{eqnarray*}
in the case $w_e<w_0$. The potential of TSF with $w_e>w_0$ is:
\begin{eqnarray*}
&&U(\xi-\xi_0)=\frac{3H_0^2}{8\pi G}\frac{w_e-w_0}{1+w_e}\left[\sin\left(\frac{3}{2}H_0\sqrt{w_e-w_0}\left(\xi-\xi_0\right)+\tan^{-1}\left(\sqrt{\frac{w_e-w_0}{1+w_0}}\right)\right)\right]^{-2}\\
&&\times\left[\sin^2\left(\frac{3}{2}H_0\sqrt{w_e-w_0}\left(\xi-\xi_0\right)+\tan^{-1}\left(\sqrt{\frac{w_e-w_0}{1+w_0}}\right)\right)\right.\\
&&\left.-w_e\cos^2\left(\frac{3}{2}H_0\sqrt{w_e-w_0}\left(\xi-\xi_0\right)+\tan^{-1}\left(\sqrt{\frac{w_e-w_0}{1+w_0}}\right)\right)\right]^{\frac{1}{2}},
\end{eqnarray*}
the corresponding potential of TSF with $w_e<w_0$ reads:
\begin{eqnarray*}
&&U(\xi-\xi_0)=\frac{3H_0^2}{8\pi G}\frac{w_0-w_e}{1+w_e}\left[\sinh\left(\frac{3}{2}H_0\sqrt{w_0-w_e}\left(\xi-\xi_0\right)+\tanh^{-1}\left(\sqrt{\frac{w_0-w_e}{1+w_0}}\right)\right)\right]^{-2}\\
&&\times\left[-\sinh^2\left(\frac{3}{2}H_0\sqrt{w_0-w_e}\left(\xi-\xi_0\right)+\tanh^{-1}\left(\sqrt{\frac{w_0-w_e}{1+w_0}}\right)\right)\right.\\
&&\left.-w_e\cosh^2\left(\frac{3}{2}H_0\sqrt{w_0-w_e}\left(\xi-\xi_0\right)+\tanh^{-1}\left(\sqrt{\frac{w_0-w_e}{1+w_0}}\right)\right)\right]^{\frac{1}{2}}.
\end{eqnarray*}
For $w_e>w_0$ these potentials have been found in \cite{Gorini2003}, for $w_e<w_0$ the corresponding potentials can be derived from them. In Fig.~\ref{fig2} the potentials of CSF and TSF are presented for the realistic multicomponent model as well as for the simple scalar field one. The potentials in both models are very different at early epoch but become similar in the epoch corresponding to the dark energy domination in the multicomponent model. In both models the potentials of CSF with $w_e<w_0$ become negative at some time in future while the potentials of TSF become imaginary. The potentials of fields with $w_e>w_0$ have no such peculiarities. It should be noted that for scalar fields with $w_e>w_0$ the field variables tend to a finite value at infinite time.

\begin{figure}
\centerline{\includegraphics[height=0.75\textheight]{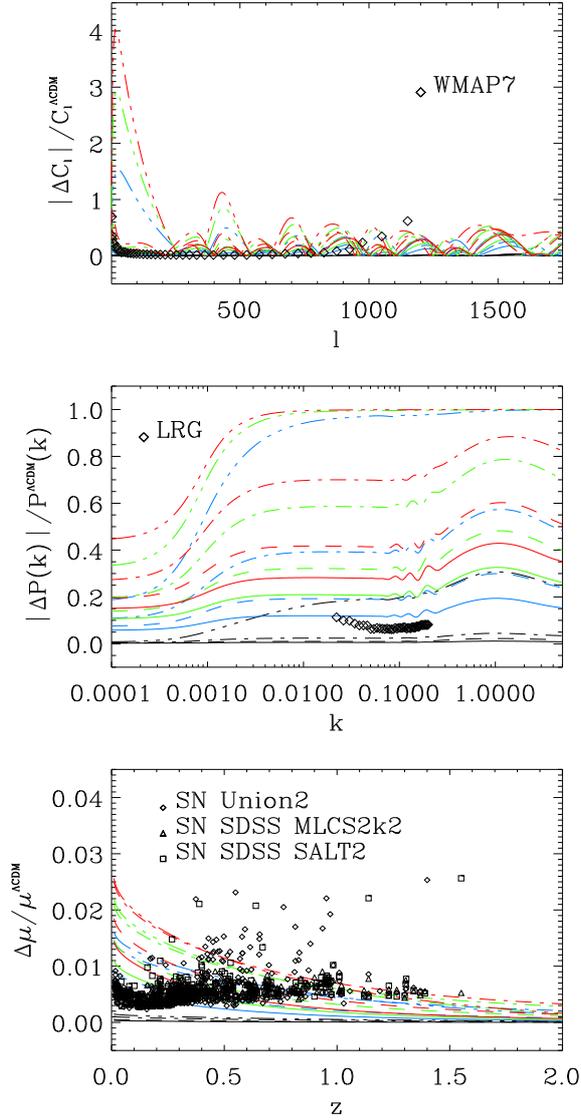}} 
\caption{From top to bottom: the relative differences of CMB temperature fluctuations power spectra, matter density power spectra and distance moduli in cosmological models with CSF with barotropic EoS as dark energy and in the fiducial $\Lambda$CDM model. Black lines correspond to $w_0=-0.99$, blue to $w_0=-0.77$, green to $w_0=-0.55$ and red to $w_0=-0.33$. Solid lines correspond to $w_e=-0.99$, dashed ones to $w_e=-0.66$, dash-dotted ones to $w_e=-0.33$ and dash-3-dotted ones to $w_e=0$.}
\label{fig3}
\end{figure}

\section{Observational constraints from current datasets}\label{curoc}

In the previous section it has been shown that the dynamics of the expansion of the Universe and its future are determined by the relation between the parameters $w_0$ and $w_e$. To find out which scenario is valid these parameters should be determined from the observational data. As the values of other cosmological parameters are unknown, the determination has to be performed for the full set of cosmological parameters which involves also the dark energy density parameter $\Omega_{de}$, the physical density parameters of baryons $\Omega_bh^2$ and cold dark matter $\Omega_{cdm}h^2$, the Hubble constant $H_0$, the spectral index of initial matter density power spectrum $n_s$, the amplitude of initial matter density power spectrum $A_s$ and the reionization optical depth $\tau$ (here and below $h=H_0/100\;km/[s\cdot Mpc]$). Instead of $H_0$ we vary the commonly used parameter $\theta$ defining the ratio of the sound horizon to the angular diameter distance multiplied by 100. These are nine unknown parameters, but the number of independent ones is 8, since we assume spatial flatness, hence the dark energy density parameter is obtained from the zero curvature condition: $\Omega_{de}=1-\Omega_b-\Omega_{cdm}$. We neglect the contribution from the tensor mode of perturbations and the masses of active neutrinos. 

\begin{figure}
\centerline{\includegraphics[width=\textwidth]{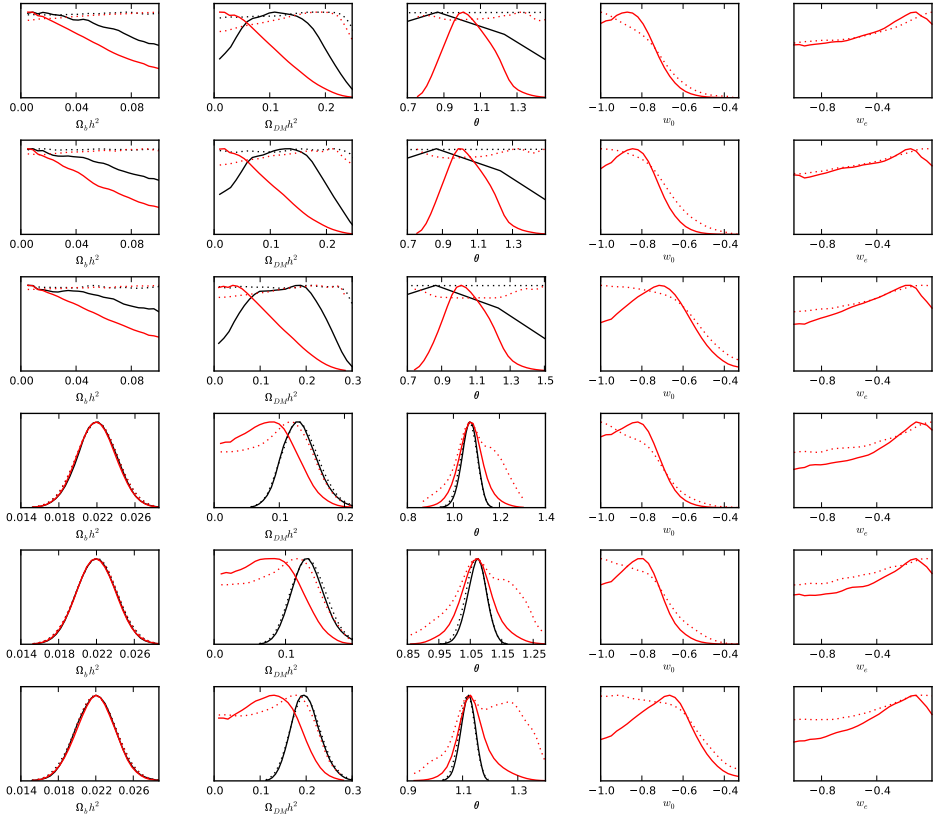}} 
\caption{One-dimensional marginalized posteriors (solid lines) and mean likelihoods (dotted lines) for the main cosmological parameters derived from the datasets SN Union2, SN SDSS SALT2, SN SDSS MLCS2k2, SN Union2+HST+BBN, SN SDSS SALT2+HST+BBN and SN SDSS MLCS2k2+HST+BBN (from top to bottom). Black lines -- $\Lambda$CDM model, red -- model with CSF.}
\label{fig5}
\end{figure}

\subsection{Method and data}

In our previous paper \cite{Novosyadlyj2010a} we have found that all cosmological parameters are determined well with the exception of $w_e$, which remains essentially unconstrained by the combined datasets used there. In order to find the best-fit value of this parameter and its confidence limits for scalar fields with both types of Lagrangians in this paper we perform the Markov Chain Monte-Carlo (MCMC) analysis for different combined datasets. 

For our analysis we use the publicly available package CosmoMC \cite{cosmomc,cosmomc_source} including the publicly available code CAMB \cite{camb,camb_source} for the calculation of the model predictions. This code has been modified to include the dark energy models proposed here as described in our previous paper~\cite{Novosyadlyj2010a}.

Each of the performed MCMC runs has 8 chains converged to $R-1<0.01$.

We use the following datasets:
\begin{itemize}
 \item \textit{CMB temperature fluctuations and polarization angular power spectra} from the 7-year WMAP observations (hereafter WMAP7) \cite{wmap7a,wmap7b};
 \item \textit{Baryon acoustic oscillations} in the space distribution of galaxies from SDSS DR7 (hereafter BAO) \cite{Percival2009};
 \item \textit{Power spectrum of luminous red galaxies} from SDSS DR7 (hereafter SDSS LRG7) \cite{Reid2009} -- in this case we obtain the nonlinear correction of the small-scale matter power spectrum using the version of \texttt{halofit} modified to include the background dynamics in models with scalar fields with barotropic EoS neglecting the dark energy perturbations which have been found to be significantly smaller than the dark matter ones at these scales at the current epoch \cite{Sergijenko2009b,Novosyadlyj2010a};
 \item \textit{Hubble constant measurements} from HST (hereafter HST) \cite{Riess2009};
 \item \textit{Big Bang Nucleosynthesis prior} on baryon abundance (hereafter BBN) \cite{bbn,Wright2007};
 \item \textit{Union2 supernovae Ia compilation} including 557 SN with SALT2 method of light curve fitting (hereafter SN Union2) \cite{SNUnion2};
 \item \textit{SDSS supernovae Ia compilation} (hereafter SN SDSS) \cite{Kessler2009} -- full sample includes 288 SN; both SALT2 \cite{Guy2007} and MLCS2k2 \cite{Jha2007} (modified by \cite{Kessler2009}) methods of light curve fitting are used. 
\end{itemize}

\subsection{The effect of scalar fields with barotropic EoS on characteristics of expansion dynamics and large scale structure of the Universe}

Let us briefly discuss the effect of the dark energy models studied here on measurable characteristics of the dynamics of expansion and large scale structure of the Universe. 
In Fig.~\ref{fig3} we present the relative differences of the CMB temperature fluctuations power spectra $(C_{\ell}-C_{\ell}^{(\Lambda CDM)})/C_{\ell}^{(\Lambda CDM)}$, matter density power spectra $(P(k)-P^{(\Lambda CDM)}(k))/P^{(\Lambda CDM)}(k)$ and distance moduli $(\mu-\mu^{(\Lambda CDM)})/\mu^{(\Lambda CDM)}$ (the distance modulus is given by the luminosity distance as $\mu=5\log_{10}D_L+25$) for models with CSF with different values of $w_0$ and $w_e$. For comparison we use the fiducial flat $\Lambda$CDM model with parameters 
$\Omega_bh^2=0.0223$, $\Omega_{cdm}h^2=0.105$, $h=0.7$, $\tau=0.09$, $n_s=0.95$ and $A_s=2.0\cdot 10^{-9}$. The main cosmological parameters in the models with a scalar field are taken to be equal to the fiducial ones. The symbols in the figure show the relative errors of the used observational data.

In the top panel we see that the dark energy with $w_0=-0.99$ leads to differences in CMB temperature fluctuations power spectrum which are at the percent level. On the other hand the dark energy with $w_e=0$ causes at low multipoles relative differences of factors $\sim 1.5-4$ for all values of $w_0$ which are not very close to $-1$. This suggests that the values of $w_e$ close to 0 should be ruled out by the CMB data with sufficiently high precision. The relative errors of observational WMAP7 data points are smaller than most of the relative differences between model spectra up to $\ell\approx1000$, however the cosmological parameters, which 
have been fixed, in Fig.~\ref{fig3} should   in practice be determined from the data jointly with the dark energy parameters $w_0$ and $w_e$. This reduces the possibility of distinguishing between the models with different values of $w_0$ and $w_e$.

The CMB allows a reliable determination of the main cosmological parameters: we have checked that the one-dimensional marginalized posteriors and mean likelihoods derived from WMAP7 only for the $\Lambda$CDM model and the model with CSF with barotropic EoS are close, which means that the proposed dark energy model does not cause the large deviations of the best-fit values of main cosmological parameters from their values in models with cosmological constant, however for $w_e$ the shapes of the marginalized posterior and mean likelihood are different, which reflects the fact that this parameter cannot be constrained using CMB data only due to the small dark energy density (comparing to the critical one) at recombination.

From the middle panel of Fig.~\ref{fig3} it follows that the dark energy with barotropic EoS also causes differences in $P(k)$  at level of percents and tens of percents for all values of $w_0$ and $w_e$ not very close to $-1$. This suggests that the data on the matter power spectrum could be useful for constraining the parameters of the studied models. Note that in the case of $P(k)$ the maximal relative differences are smaller than for the CMB and their absolute value never exceeded 1. Similarly to the CMB spectrum from WMAP7, for values of $w_0$ not very close to $-1$ the relative errors of the SDSS LRG7 data are smaller than the relative differences of power spectra between models with CSF with barotropic EoS and $\Lambda$CDM. 

The data on the power spectrum of luminous red galaxies should be used combined with CMB datasets or priors, in such combinations they serve as an additional constraints on the 
parameters of the scalar fields with barotropic EoS and other cosmological parameters.

In the bottom panel the relative differences of the distance moduli are shown. They decrease with redshift while the distance modulus and luminosity distance grow. These differences are the smallest among all studied, they do not exceed 0.03, but up to $z\sim0.5$ they are larger than most of the relative errors of supernovae distance moduli from used compilations and up to $z\sim1$ they remain comparable to these relative errors of data for all values of the current scalar field EoS parameter that are not very close to $-1$. The supernovae data are known to be the key data for determination of the best-fit values of cosmological parameters and their confidence range in models with dark energy with variable EoS. At the beginning of the next subsection we shall check their importance for constraining our particular model.

Other data providing distance information are also BAO and HST. BAO should be used in combination with WMAP7 or with additional priors on parameters derived from the CMB datasets, in these combinations they can provide additional constraints on values of all cosmological parameters. HST data are useful for the determination of $H_0$ as well as to provide an additional constraint on deviations of cosmological models from the flat $\Lambda$CDM model with $\Omega_{\Lambda}=0.7$ and $H_0=74.2\;km/[s\cdot Mpc]$ at $z=0.04$ when combined with other datasets. We also use the BBN prior on baryon abundance that  puts an additional constraint on $\Omega_bh^2$.

In this paper we use flat priors for all parameters, for $w_0$ and $w_e$ the allowed ranges are set to $-1\leq w_0\leq-0.33$ and $-1\leq w_e\leq0$. The reason for the lower bound on both parameters is that we study the quintessential scalar fields, for which the field variables become imaginary if the EoS parameter is smaller than $-1$. The upper bound on $w_0$ corresponds to the value at which the accelerated expansion of the Universe at current epoch is no longer possible. The upper bound on the early EoS parameter can be obtained using e.g. Fig.~\ref{fig3}, in the top panel of which we see that $w_e=0$ can cause so large changes in the CMB power spectrum that larger values of $w_e$ should be avoided.

\begin{figure}
\centerline{\includegraphics[width=\textwidth]{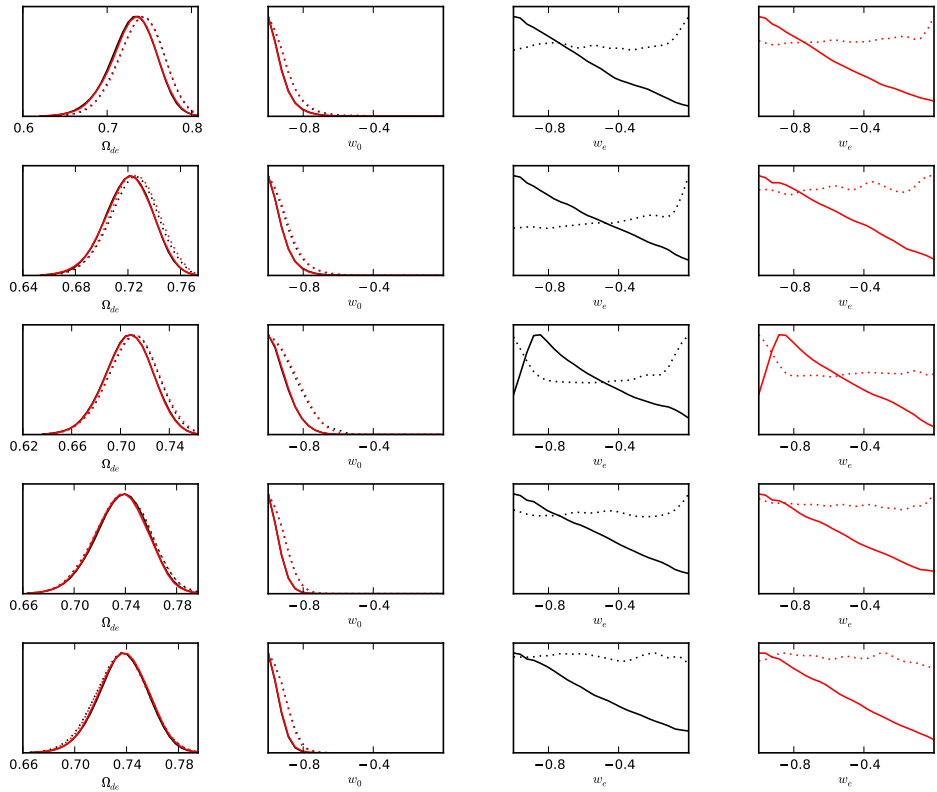}} 
\caption{One-dimensional marginalized posteriors (solid lines) and mean likelihoods (dotted lines) for the datasets WMAP7+HST+BBN, WMAP7+HST+BBN+BAO, WMAP7+HST+BBN+SDSS LRG7, WMAP7+HST+BBN+SN Union2 and WMAP7+HST+BBN+SN SDSS SALT2 (from top to bottom). From left to right: the functions for the dark energy density $\Omega_{de}$, the current value of dark energy EoS parameter $w_0$ for both CSF (black lines) and TSF (red), the early value of dark energy EoS parameter $w_e$ for CSF and the early value of dark energy EoS parameter $w_e$ for CSF.}
\label{fig6}
\end{figure}

\begin{figure}
\centerline{\includegraphics[width=0.65\textwidth]{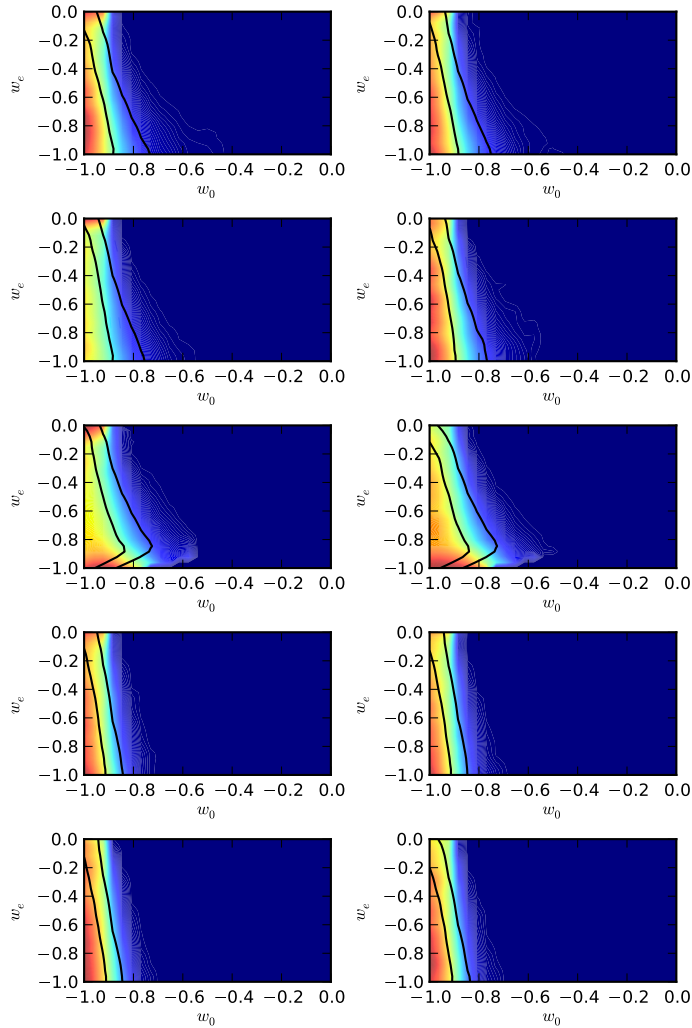}} 
\caption{Two-dimensional mean likelihood distributions in the plane $w_e-w_0$ for the same datasets and models as in Fig.~\ref{fig6}. Solid lines show the $1\sigma$ and $2\sigma$ confidence contours.}
\label{fig7}
\end{figure}

\subsection{Results}

At the beginning we analyse the possibility of constraining the cosmological parameters of models with scalar fields with barotropic EoS as dark energy using the supernovae data.

In 3 upper rows of Fig.~\ref{fig5} we present the results of determination of the main cosmological parameters along with $w_0$ and $w_e$ from the supernovae data only. We see that for the parameters defining the barotropic EoS, $w_0$ and $w_e$, the one-dimensional marginalized posteriors and mean likelihoods are relatively close. This means that the supernovae data indeed are important for constraining the studied dark energy models. However, the parameters $\Omega_bh^2$, $\Omega_{cdm}h^2$ and $\theta$ remain essentially unconstrained for both $\Lambda$CDM model and model with barotropic EoS. In 3 lower rows of Fig.~\ref{fig5} we present the one-dimensional marginalized posteriors and mean likelihoods for the same models and supernovae datasets combined with BBN and HST. We see that now the parameters of $\Lambda$CDM model are constrained well, while for the model with scalar field $\Omega_bh^2$ is constrained as well as in the $\Lambda$CDM case but $\Omega_{cdm}h^2$ and $\theta$ are constrained better than by supernovae data only though not perfectly. All 3 supernovae datasets provide the constraints with comparable level of accuracy, however it is worth to note that SN SDSS with MLCS2k2 fitting puts the clearly weakest constraint on $w_0$.

Now we analyse the constraints on studied dark energy models from the combined datasets including WMAP7 data. In the first step we analyze 6 combined datasets:
\begin{itemize}
 \item WMAP7+HST+BBN,
 \item WMAP7+HST+BBN+BAO,
 \item WMAP7+HST+BBN+SDSS LRG7,
 \item WMAP7+HST+BBN+SN Union2,
 \item WMAP7+HST+BBN+SN SDSS SALT2 and
 \item WMAP7+HST+BBN+SN SDSS MLCS2k2.
\end{itemize}

In Fig.~\ref{fig6} we see that for the datasets WMAP7+HST+BBN, WMAP7+HST+BBN\linebreak+BAO, WMAP7+HST+BBN+SDSS LRG7, WMAP7+HST+BBN+SN Union2 and \linebreak WMAP7+HST+BBN+SN SDSS SALT2 the shapes of the marginalized posterior distributions and the mean likelihoods for the early value of EoS parameter are different, this indicates a significant non-Gaussianity of the likelihood function for $w_e$. The difference between two-dimensional marginalized posteriors and mean likelihoods in the plane $w_e-w_0$, which is shown in Fig.~\ref{fig7}, confirms this conclusion. As it has been shown in our paper \cite{Novosyadlyj2010c} this is also the case for the combination WMAP7+HST+BBN+SN Union with the light curves of supernovae from the Union compilation \cite{Kowalski2008} fitted using the SALT method \cite{Guy2005}.

The best-fit values of the cosmological parameters (obtained from the best-fit sample) and their $1\sigma$ limits from the extremal values of the N-dimensional distribution are presented in Tables \ref{tab2}-\ref{tab3} for the combined datasets WMAP7+HST+BBN, WMAP7+HST+BBN\linebreak+BAO, WMAP7+HST+BBN+SDSS LRG7, WMAP7+HST+BBN+SN Union2 and \linebreak WMAP7+HST+BBN+SN SDSS SALT2.

\begin{table}
 \begin{tabular}{|c|c|c|c|c|c|}
 \hline
 & & & & &\\
 Parameters&CSF&CSF&CSF&
 CSF&CSF\\
 & & & & &\\
 \cline{2-6}
 & & & & &\\
 &WMAP7&WMAP7&WMAP7&WMAP7&WMAP7\\
 &&BAO&SDSS LRG7&SN Union2&SN SDSS\\
& & & & &SALT2\\& & & & &\\
 \hline& & & & &\\
$\Omega_{de}$& 0.75$_{- 0.09}^{+ 0.05}$& 0.72$_{- 0.05}^{+ 0.04}$& 0.72$_{- 0.07}^{+ 0.03}$& 0.74$_{- 0.07}^{+ 0.04}$& 0.73$_{- 0.05}^{+ 0.05}$\\& & & & &\\
$w_0$&-0.99$_{- 0.01}^{+ 0.30}$&-0.99$_{- 0.01}^{+ 0.29}$&-0.94$_{- 0.06}^{+ 0.30}$&-1.00$_{- 0.00}^{+ 0.19}$&-1.00$_{- 0.00}^{+ 0.17}$\\& & & & &\\
$w_e$&-0.92$_{- 0.08}^{+ 0.92}$&-0.01$_{- 0.99}^{+ 0.01}$&-0.97$_{- 0.03}^{+ 0.97}$&-0.49$_{- 0.51}^{+ 0.49}$&-0.72$_{- 0.28}^{+ 0.72}$\\& & & & &\\
$100\Omega_b h^2$& 2.26$_{- 0.14}^{+ 0.17}$& 2.26$_{- 0.14}^{+ 0.15}$& 2.26$_{- 0.13}^{+ 0.16}$& 2.25$_{- 0.14}^{+ 0.17}$& 2.25$_{- 0.13}^{+ 0.17}$\\& & & & &\\
$10\Omega_{cdm} h^2$& 1.08$_{- 0.12}^{+ 0.16}$& 1.11$_{- 0.13}^{+ 0.12}$& 1.11$_{- 0.12}^{+ 0.13}$& 1.09$_{- 0.13}^{+ 0.13}$& 1.11$_{- 0.15}^{+ 0.10}$\\& & & & &\\
$H_0$& 72.1$_{-  8.8}^{+  5.0}$& 69.4$_{-  5.2}^{+  4.4}$& 69.5$_{-  6.9}^{+  3.6}$& 71.6$_{-  6.1}^{+  4.8}$& 70.7$_{-  4.9}^{+  5.0}$\\& & & & &\\
$n_s$& 0.97$_{- 0.03}^{+ 0.04}$& 0.98$_{- 0.04}^{+ 0.03}$& 0.97$_{- 0.03}^{+ 0.05}$& 0.97$_{- 0.03}^{+ 0.04}$& 0.97$_{- 0.03}^{+ 0.04}$\\& & & & &\\
$\log(10^{10}A_s)$& 3.06$_{- 0.09}^{+ 0.11}$& 3.08$_{- 0.10}^{+ 0.09}$& 3.08$_{- 0.08}^{+ 0.10}$& 3.08$_{- 0.10}^{+ 0.09}$& 3.08$_{- 0.10}^{+ 0.10}$\\& & & & &\\
$z_{rei}$& 10.7$_{-  3.4}^{+  3.1}$& 10.8$_{-  3.5}^{+  2.9}$& 10.4$_{-  3.0}^{+  3.4}$& 10.8$_{-  3.5}^{+  2.9}$& 10.6$_{-  3.4}^{+  3.2}$\\& & & & &\\
$t_0$& 13.7$_{-  0.3}^{+  0.4}$& 13.8$_{-  0.4}^{+  0.3}$& 13.7$_{-  0.3}^{+  0.5}$& 13.7$_{-  0.3}^{+  0.4}$& 13.8$_{-  0.3}^{+  0.4}$\\& & & & &\\
 \hline
& & & & &\\
$-\log L$&3737.64&3738.71&3761.85&4003.14&3867.14\\
& & & & &\\
\hline
 \end{tabular}
\caption{The best-fit values for cosmological parameters and the $1\sigma$ limits from the extremal values of the N-dimensional distribution determined for the case of CSF by the MCMC technique from the combined datasets WMAP7+HST+BBN, WMAP7+HST+BBN+BAO, WMAP7+HST+BBN+SDSS LRG7, WMAP7+HST+BBN+SN Union2 and WMAP7+HST+BBN+SN SDSS SALT2.}\label{tab2}
\end{table}

\begin{table}
 \begin{tabular}{|c|c|c|c|c|c|}
 \hline
 & & & & &\\
 Parameters&TSF&TSF&TSF&
 TSF&TSF\\
 & & & & &\\
 \cline{2-6}
 & & & & &\\
 &WMAP7&WMAP7&WMAP7&WMAP7&WMAP7\\
 &&BAO&SDSS LRG7&SN Union2&SN SDSS\\
& & & & &SALT2\\& & & & &\\
 \hline& & & & &\\
$\Omega_{de}$& 0.75$_{- 0.09}^{+ 0.05}$& 0.72$_{- 0.05}^{+ 0.04}$& 0.71$_{- 0.05}^{+ 0.05}$& 0.74$_{- 0.06}^{+ 0.05}$& 0.74$_{- 0.06}^{+ 0.04}$\\& & & & &\\
$w_0$&-0.99$_{- 0.01}^{+ 0.34}$&-0.99$_{- 0.01}^{+ 0.30}$&-0.93$_{- 0.07}^{+ 0.26}$&-0.99$_{- 0.01}^{+ 0.18}$&-1.00$_{- 0.00}^{+ 0.17}$\\& & & & &\\
$w_e$&-0.54$_{- 0.46}^{+ 0.54}$&-0.77$_{- 0.23}^{+ 0.77}$&-0.96$_{- 0.04}^{+ 0.96}$&-0.76$_{- 0.24}^{+ 0.76}$&-0.22$_{- 0.78}^{+ 0.22}$\\& & & & &\\
$100\Omega_b h^2$& 2.26$_{- 0.14}^{+ 0.16}$& 2.27$_{- 0.14}^{+ 0.15}$& 2.27$_{- 0.14}^{+ 0.15}$& 2.27$_{- 0.15}^{+ 0.14}$& 2.27$_{- 0.15}^{+ 0.15}$\\& & & & &\\
$10\Omega_{cdm} h^2$& 1.07$_{- 0.12}^{+ 0.15}$& 1.13$_{- 0.15}^{+ 0.09}$& 1.13$_{- 0.12}^{+ 0.11}$& 1.09$_{- 0.13}^{+ 0.12}$& 1.09$_{- 0.13}^{+ 0.12}$\\& & & & &\\
$H_0$& 72.3$_{-  9.0}^{+  5.6}$& 69.8$_{-  5.8}^{+  4.1}$& 68.2$_{-  5.6}^{+  5.2}$& 71.4$_{-  5.9}^{+  4.7}$& 71.7$_{-  5.7}^{+  4.0}$\\& & & & &\\
$n_s$& 0.97$_{- 0.04}^{+ 0.04}$& 0.97$_{- 0.03}^{+ 0.04}$& 0.97$_{- 0.03}^{+ 0.04}$& 0.97$_{- 0.04}^{+ 0.04}$& 0.97$_{- 0.03}^{+ 0.04}$\\& & & & &\\
$\log(10^{10}A_s)$& 3.07$_{- 0.10}^{+ 0.10}$& 3.09$_{- 0.10}^{+ 0.08}$& 3.09$_{- 0.09}^{+ 0.09}$& 3.07$_{- 0.09}^{+ 0.11}$& 3.06$_{- 0.08}^{+ 0.11}$\\& & & & &\\
$z_{rei}$& 10.8$_{-  3.6}^{+  3.2}$& 10.6$_{-  3.4}^{+  3.1}$& 10.7$_{-  3.5}^{+  3.1}$& 10.3$_{-  3.0}^{+  3.4}$& 10.3$_{-  3.0}^{+  3.5}$\\& & & & &\\
$t_0$& 13.7$_{-  0.3}^{+  0.4}$& 13.8$_{-  0.3}^{+  0.4}$& 13.8$_{-  0.3}^{+  0.4}$& 13.7$_{-  0.3}^{+  0.4}$& 13.7$_{-  0.3}^{+  0.4}$\\& & & & &\\
 \hline
& & & & &\\
$-\log L$&3737.69&3738.92&3762.10&4003.12&3867.19\\
& & & & &\\
\hline
 \end{tabular}
\caption{The best-fit values for cosmological parameters and the $1\sigma$ limits from the extremal values of the N-dimensional distribution determined for the case of TSF by the MCMC technique from the combined datasets WMAP7+HST+BBN, WMAP7+HST+BBN+BAO, WMAP7+HST+BBN+SDSS LRG7, WMAP7+HST+BBN+SN Union2 and WMAP7+HST+BBN+SN SDSS SALT2.}\label{tab3}
\end{table}

\begin{figure}
\centerline{\includegraphics[width=\textwidth]{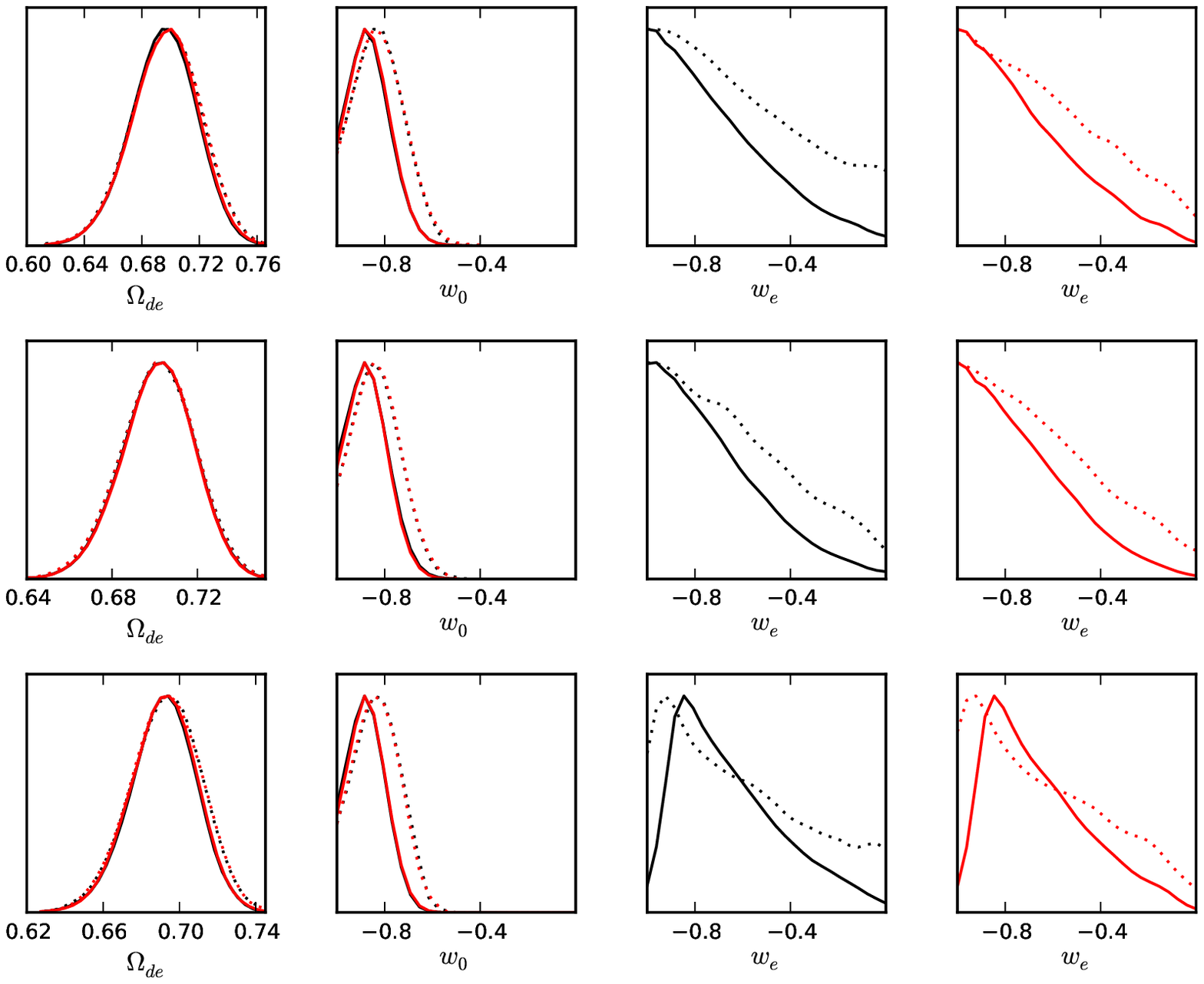}} 
\caption{One-dimensional marginalized posteriors (solid lines) and mean likelihoods (dotted ones) for the combined datasets WMAP7+HST+BBN+SN SDSS MLCS2k2, WMAP7+HST+BBN+SN SDSS MLCS2k2+BAO and WMAP7+HST+BBN+SN SDSS MLCS2k2+SDSS LRG7 (from top to bottom). From left to right: the functions for the dark energy density $\Omega_{de}$, the current value of dark energy EoS parameter $w_0$ for both CSF (black lines) and TSF (red), the early value of dark energy EoS parameter $w_e$ for CSF and the early value of dark energy EoS parameter $w_e$ for CSF.}
\label{fig8}
\end{figure}

\begin{figure}
\centerline{\includegraphics[width=0.65\textwidth]{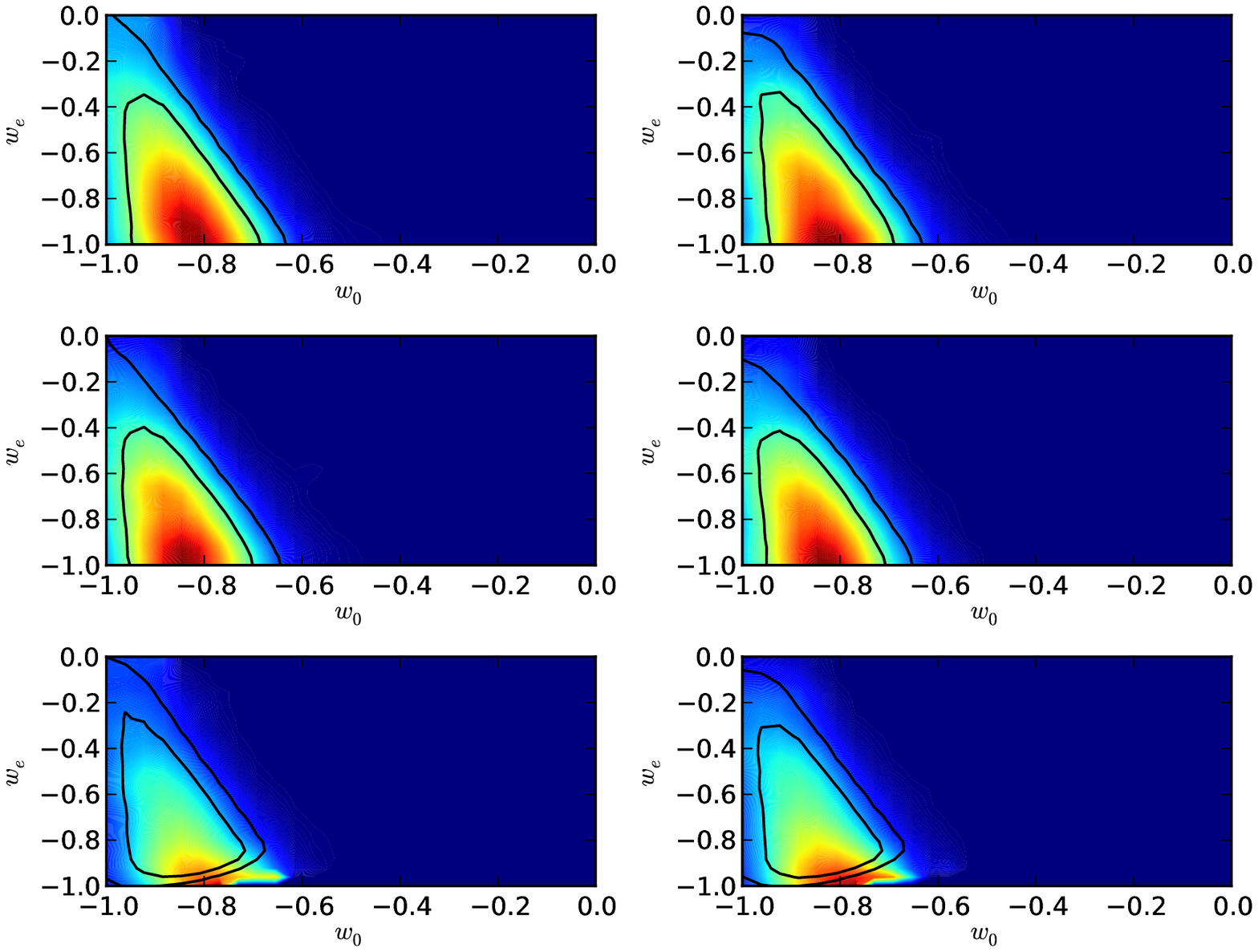}} 
\caption{Two-dimensional mean likelihood distributions in the plane $w_e-w_0$  for corresponding datasets and models from Fig.~\ref{fig8}. Solid lines show the $1\sigma$ and $2\sigma$ confidence contours.}
\label{fig9}
\end{figure}

The situation is different for the dataset WMAP7+HST+BBN+SN SDSS MLCS2k2. In this case, as it can be seen in the top panels of Fig.~\ref{fig8}, the shapes of mean likelihoods and the one-dimensional marginalized posteriors for $w_e$ are similar to the half-Gaussian with center at the boundary of the allowed range of values, $w_e=-1$. This dataset can be used for a reliable estimation of the early EoS parameter. The difference between both curves is a signal of non-Gaussianity, which is however substantially reduced compared to all other datasets. The two-dimensional mean likelihoods and marginalized posteriors in the plane $w_e-w_0$ presented in the top panels of Fig.~\ref{fig9} support this conclusion. The shapes of the high-likelihood regions are similar to the shapes of $1\sigma$ and $2\sigma$ confidence contours.

Hence, we have found that the data on SN Ia from the SDSS compilation with modified MLCS2k2 fitting of light curves allow to constrain $w_e$ while the same data with SALT2 fitting do not. This is a demonstration of the well-known discrepancy between SALT2 and MLCS2k2  which is due mainly to the different rest-frame U-band models and the assumptions about the color variations in both fitting methods \cite{Kessler2009}. 

\begin{figure}
\centerline{\includegraphics[width=\textwidth]{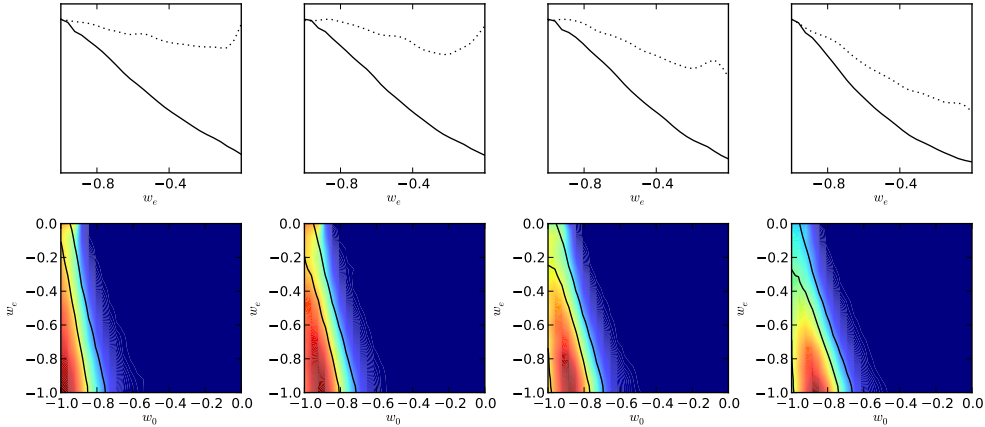}} 
\caption{Top: one-dimensional marginalized posteriors (solid lines) and mean likelihoods (dotted ones) for CSF and the combined datasets WMAP7+HST+BBN+SN SDSS MLCS2k2 for the subsamples NEARBY+SDSS+ESSENCE, NEARBY+SDSS+HST, NEARBY+SDSS+ESSENCE+HST and NEARBY+SDSS+SNLS of SN SDSS compilation (from left to right). Bottom: corresponding two-dimensional mean likelihood distributions in the plane $w_e-w_0$. Solid lines show the $1\sigma$ and $2\sigma$ confidence contours.}
\label{fig11}
\end{figure}

In paper \cite{Novosyadlyj2010c} we have performed similar MCMC runs for the combined datasets including the SN subset NEARBY+SDSS (136 SN) from the SDSS compilation, for which this discrepancy is smallest~\cite{Kessler2009}. In this case the parameter $w_e$ remains unconstrained for SN data with both light curve fitting methods. Therefore, the non-Gaussianity of the likelihood function with respect to $w_e$ is reduced by inclusion of the higher-redshift SN samples, for which the treatment of the MLCS2k2 method differs significantly from the corresponding treatment of the SALT2 method. In Fig.~\ref{fig11} we present the one- and two-dimensional marginalized posteriors and mean likelihoods derived from the MCMC runs performed for the case of CSF for combined datasets including WMAP7, HST, BBN, NEARBY+SDSS SN data and in addition different higher-redshift subsamples of the SDSS compilation with MLCS2k2 light curve fitting. We see that the reduction of the non-Gaussianity is due mainly to the inclusion of the SNLS subsample, for which the difference between SALT2 and MLCS2k2 is found to increase with redshift \cite{Kessler2009}.

\begin{figure}
\centerline{\includegraphics[width=0.9\textwidth]{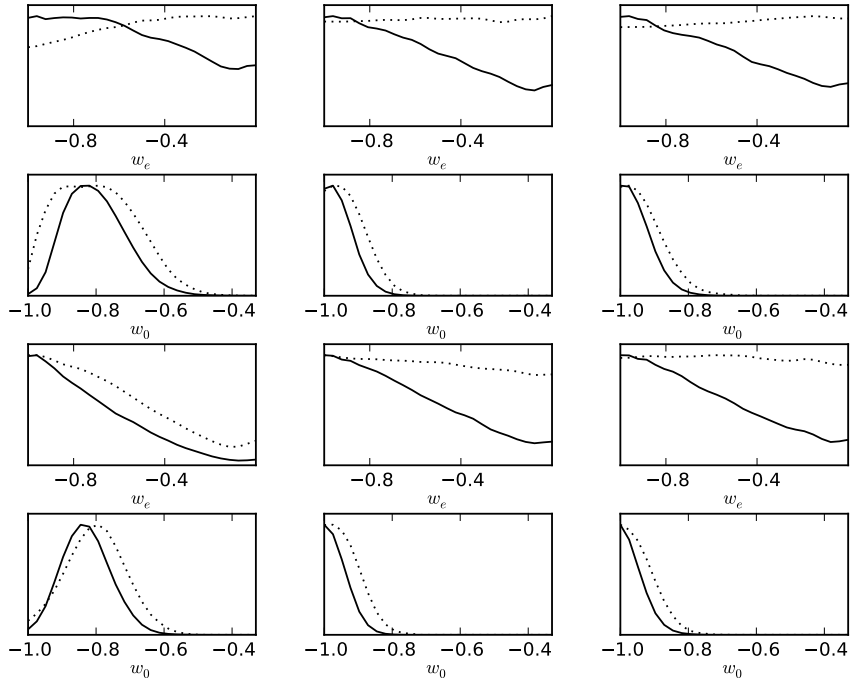}} 
\caption{One-dimensional marginalized posteriors (solid lines) and mean likelihoods (dotted lines) for the datasets SN SDSS MLCS2k2, SN SDSS SALT2, SN Union2 (2 upper rows, from left to right) and SN SDSS MLCS2k2+HST+BBN, SN SDSS SALT2+HST+BBN, SN Union2+HST+BBN (2 lower rows, from left to right) for the models where only $w_0$ and $w_e$ are allowed to vary.}
\label{fig10}
\end{figure}

Let us fix the parameters $\Omega_bh^2$, $\Omega_{cdm}h^2$ and $\theta$ to their fiducial values and determine the constraints on $w_0$ and $w_e$ from the supernovae data only. In the 2 upper rows of Fig.~\ref{fig10} the one-dimensional marginalized posteriors and mean likelihoods are shown for $w_e$ and $w_0$. We see that in this case none of the supernovae data allow a reliable determination of $w_e$. The constraints on $w_0$ obtained from SN SDSS MLCS2k2 are also weak. Now we combine the supernovae data with the HST ones. In the 2 bottom rows of Fig.~\ref{fig10} the corresponding the one-dimensional marginalized posteriors and mean likelihoods for $w_e$ and $w_0$ are presented. We find that the SN SDSS SALT2 and SN Union2 data still do not allow to put constraints on $w_e$ while in the case of SN SDSS data the marginalized posterior and mean likelihood are close. Thus the SN SDSS MLCS2k2 data constrain the values of $w_e$ when combined with the additional constraint on the deviations of model from $\Lambda$CDM coming from HST. The transition of positions of the peaks of posterior and likelihood from values close to 0 in Fig.~\ref{fig5} to $-1$ is due to the fact that the parameters $\Omega_{cdm}h^2$ and $\theta$ which are highly correlated with $w_e$ and $w_0$ in the case presented in Fig.~\ref{fig5} are now fixed to their fiducial values. The current value of the dark energy EoS parameter $w_0$ is constrained well by all 3 supernovae datasets combined with HST. While SN SDSS SALT2 and SN Union2 prefer values close to $-1$, the best-fit values of $w_0$ obtained from SN SDSS MLCS2k2 are closer to $-0.8$.

A general discussion of the differences, benefits and limitations of the SALT2 and MLCS2k2 light curve fitting methods is beyond the scope of this paper. Here we should mention only that the MLCS2k2 is believed to be more model-independent than SALT2 which fits the global parameters $M$, $\alpha$ and $\beta$ determining the distance moduli together with the cosmological ones. Let us compare the root mean squares of errors of different supernovae samples. For SN Union2 it is $0.269$, for SN SDSS SALT2 it equals $0.260$ while for SN SDSS MLCS2k2 $0.233$. Thus the supernovae from the SDSS compilation with MLCS2k2 fitting have the smallest errors among all used compilations. These small difference results in the possibility of reliable determination of the best-fit value of $w_e$ and its confidence ranges. Considering the subsamples of the SN SDSS compilation, the datasets NEARBY+SDSS SN have the root mean square of errors $0.198$ for MLCS2k2 and $0.191$ for SALT2 -- these values are almost the same, so there is no difference in possibility of constraining $w_e$ from these data. Among the higher-redshift subsamples of SN SDSS compilation the samples ESSENCE, HST and SNLS have the root mean squares of errors $0.261$, $0.283$ and $0.246$ correspondingly. Thus, the inclusion of SNLS subsample is crucial for constraining of the early value of dark energy EoS parameter since it has the smallest errors among all higher-redshift subsamples of SN SDSS compilation with MLCS2k2 light curve fitting method.

As the next step we perform the similar MCMC runs for two combined datasets including WMAP7+HST+BBN+SN SDSS MLCS2k2 and the data on large-scale structure:
\begin{itemize}
 \item WMAP7+HST+BBN+SN SDSS MLCS2k2+BAO and
 \item WMAP7+HST+BBN+SN SDSS MLCS2k2+SDSS LRG7.
\end{itemize}
The one- and two-dimensional marginalized posteriors and mean likelihoods for these datasets are shown in middle and bottom panels of Fig.~\ref{fig8}-\ref{fig9}. These datasets allow the constraints on the value of $w_e$ at comparable level of accuracy as the previous dataset WMAP7+HST+BBN\linebreak+SN SDSS MLCS2k2. For the dataset 
WMAP7+HST+BBN+SN SDSS MLCS2k2+BAO the one-dimensional marginalized posterior and mean likelihood have the shape of the half-Gaussian with center at the boundary of the allowed range of values ($w_e=-1$). The small difference between both curves indicates the slight non-Gaussianity of the likelihood function with respect to $w_e$, which however does not reduce the possibility of a reliable estimation of $w_e$ from these data. For the set WMAP7+HST+BBN+SN SDSS MLCS2k2+SDSS LRG7 they have the shape of the Gaussians with centers at slightly different and larger values of $w_e$. This difference signals that some non-Gaussianity of the likelihood function with respect to the early EoS parameter exists, however, as it is relatively small, we conclude that the last dataset can also be used for the reliable estimation of $w_e$.

The two-dimensional $w_e-w_0$ marginalized posteriors and mean likelihoods presented in the middle and bottom panels of Fig.~\ref{fig9}  support these conclusions. For the dataset WMAP7+HST+BBN+SN SDSS MLCS2k2+BAO the shapes of the high-likelihood regions and of the $1\sigma$ and $2\sigma$ confidence contours are similar. For the combination WMAP7+HST\linebreak+BBN+SN SDSS MLCS2k2+SDSS LRG7 the high-likelihood region lays partially outside the $2\sigma$ confidence contour, this is a signal of the above mentioned non-Gaussianity, which can be however neglected since it is relatively small.

From Fig.~\ref{fig8} we see that the fields mimicking a cosmological constant are excluded at the $1\sigma$ confidence level for all 3 datasets for both Lagrangians. Moreover, for the set WMAP7+HST+BBN+SN SDSS MLCS2k2+SDSS LRG7 both such fields lay even slightly outside the $2\sigma$ confidence contour. This is due to the inclusion of SN SDSS data with light curves fitted using the modified MLCS2k2 method. The values of $w_e$ close to 0 are excluded nearly at the $2\sigma$ confidence level for both fields and all 3 datasets.

The best-fit values of the cosmological parameters and their $1\sigma$ limits from the extremal values of the N-dimensional distribution are presented in Table \ref{tab1} for CSF and TSF models with barotropic EoS for the combined datasets WMAP7+HST+BBN+SN SDSS MLCS2k2, WMAP7+HST+BBN+SN SDSS MLCS2k2+BAO and WMAP7+HST+BBN\linebreak+SN SDSS MLCS2k2+SDSS LRG7. Note that these limits are significantly wider than the corresponding limits obtained from the one- and two-dimensional marginalized distributions. We see that for all cases including SN SDSS MLCS2k2 the best-fit model has $w_e<w_0$, thus the repulsive character of the scalar fields will stop and the expansion of such Universe will turn into collapse.

It should be noted that the fields with classical and tachyonic Lagrangians cannot be distinguished by the currently available data: the differences between the best-fit parameters are within the corresponding $1\sigma$ confidence limits (see also \cite{Novosyadlyj2010b}). Moreover, as we can see in the bottom rows of Tables \ref{tab2}-\ref{tab1}, the values of $-\log L$ differ by less than 1 for the best-fit parameters of models with CSF and TSF obtained from the same datasets. Note that the values of $-\log L$ for the WMAP7+HST+BBN+SN SDSS SALT2 dataset are higher than those for the same dataset including SN SDSS with MLCS2k2 light curve fitting method. 

It should be noted that, as it can be seen in Fig.~\ref{fig8} and \ref{fig9}, the dark energy density $\Omega_{de}$ and its current value of EoS parameter $w_0$ are well constrained by all the used combined datasets. The curves for CSF and TSF overlap.

Finally, let us discuss the best-fit values of Hubble constant obtained from different datasets. As it can be seen in Tables \ref{tab2}-\ref{tab1}, for the combined datasets, which do not include SN SDSS MLCS2k2 data, the best-fit values of $H_0$ are in the range 68.2-72.3 $km/(s\cdot Mpc)$ which is closer to $H_0=74.2\;km/(s\cdot Mpc)$ from \cite{Riess2009} than the range 65.9-67.1 $km/(s\cdot Mpc)$ obtained when we include these data. Performing the MCMC runs for the model with CSF and the combined datasets including WMAP7, HST, BBN and different subsamples of SN SDSS compilation with MLCS2k2 light curve fitting we have found the following fit values of Hubble constant (in $km/(s\cdot Mpc)$): 71.8 for NEARBY+SDSS SN (and 71.5 for the SALT2 fitting method), 70.1 for NEARBY+SDSS+ESSENCE SN, 68.5 for NEARBY+SDSS+HST SN, 67.7 for NEARBY+SDSS+SNLS SN and 67.5 for NEARBY\linebreak+SDSS+ESSENCE+HST SN. Thus, in the case of MLCS2k2 fitting of light curves the best-fit value of $H_0$ is lowered mainly by the subsamples either ESSENCE+HST or SNLS from the SN SDSS compilation.

\begin{table}
 \begin{tabular}{|c|c|c|c|c|c|c|}
 \hline
 & & & & & &\\
 Parameters&CSF&CSF&CSF&
 TSF&TSF&TSF\\
 & & & & & &\\
 \cline{2-7}
 & & & & & &\\
 &WMAP7&WMAP7&WMAP7&WMAP7&WMAP7&WMAP7\\
  &SN SDSS&SN SDSS&SN SDSS&SN SDSS&SN SDSS&SN SDSS\\
 &MLCS2k2&MLCS2k2&MLCS2k2&MLCS2k2&MLCS2k2&MLCS2k2\\
 &&BAO&SDSS LRG7&&BAO&SDSS LRG7\\
 & & & & & &\\
 \hline
 & & & & & &\\
$\Omega_{de}$& 0.70$_{- 0.07}^{+ 0.06}$& 0.71$_{- 0.05}^{+ 0.04}$& 0.69$_{- 0.05}^{+ 0.04}$& 0.70$_{- 0.07}^{+ 0.06}$& 0.70$_{- 0.05}^{+ 0.04}$& 0.69$_{- 0.05}^{+ 0.05}$\\
 & & & & & &\\
$w_0$&-0.81$_{- 0.19}^{+ 0.20}$&-0.85$_{- 0.15}^{+ 0.23}$&-0.84$_{- 0.16}^{+ 0.19}$&-0.81$_{- 0.19}^{+ 0.24}$&-0.85$_{- 0.15}^{+ 0.23}$&-0.84$_{- 0.16}^{+ 0.18}$\\
 & & & & & &\\
$w_e$&-1.00$_{- 0.00}^{+ 0.99}$&-0.98$_{- 0.02}^{+ 0.98}$&-0.93$_{- 0.07}^{+ 0.92}$&-0.98$_{- 0.02}^{+ 0.96}$&-0.99$_{- 0.01}^{+ 0.98}$&-0.93$_{- 0.06}^{+ 0.91}$\\
 & & & & & &\\
$100\Omega_b h^2$& 2.27$_{- 0.14}^{+ 0.17}$& 2.28$_{- 0.15}^{+ 0.15}$& 2.28$_{- 0.15}^{+ 0.16}$& 2.31$_{- 0.18}^{+ 0.13}$& 2.26$_{- 0.13}^{+ 0.17}$& 2.26$_{- 0.13}^{+ 0.18}$\\
 & & & & & &\\
$10\Omega_{cdm} h^2$& 1.09$_{- 0.15}^{+ 0.17}$& 1.10$_{- 0.14}^{+ 0.13}$& 1.12$_{- 0.14}^{+ 0.12}$& 1.10$_{- 0.15}^{+ 0.15}$& 1.10$_{- 0.13}^{+ 0.13}$& 1.11$_{- 0.13}^{+ 0.12}$\\
 & & & & & &\\
$H_0$& 66.0$_{-  5.1}^{+  5.5}$& 67.1$_{-  4.4}^{+  3.8}$& 66.2$_{-  4.4}^{+  3.8}$& 66.5$_{-  5.4}^{+  4.9}$& 66.7$_{-  4.1}^{+  4.2}$& 65.9$_{-  4.3}^{+  4.2}$\\
 & & & & & &\\
$n_s$& 0.97$_{- 0.04}^{+ 0.04}$& 0.98$_{- 0.04}^{+ 0.04}$& 0.97$_{- 0.03}^{+ 0.04}$& 0.97$_{- 0.04}^{+ 0.04}$& 0.97$_{- 0.03}^{+ 0.04}$& 0.98$_{- 0.04}^{+ 0.04}$\\
 & & & & & &\\
$\log(10^{10}A_s)$& 3.07$_{- 0.08}^{+ 0.11}$& 3.09$_{- 0.11}^{+ 0.09}$& 3.10$_{- 0.11}^{+ 0.09}$& 3.08$_{- 0.10}^{+ 0.10}$& 3.07$_{- 0.09}^{+ 0.11}$& 3.08$_{- 0.09}^{+ 0.10}$\\
 & & & & & &\\
$z_{rei}$& 10.3$_{-  3.1}^{+  3.7}$& 11.0$_{-  3.8}^{+  2.9}$& 10.8$_{-  3.5}^{+  3.0}$& 10.8$_{-  3.4}^{+  3.1}$& 10.3$_{-  2.9}^{+  3.4}$& 10.3$_{-  2.9}^{+  3.5}$\\
 & & & & & &\\
$t_0$& 13.8$_{-  0.3}^{+  0.5}$& 13.8$_{-  0.3}^{+  0.4}$& 13.8$_{-  0.3}^{+  0.5}$& 13.8$_{-  0.3}^{+  0.5}$& 13.8$_{-  0.3}^{+  0.4}$& 13.8$_{-  0.3}^{+  0.4}$\\
 & & & & & &\\
 \hline
& & & & & &\\
$-\log L$&3859.27&3860.38&3882.28&3859.31&3860.28&3882.37\\
& & & & & &\\
\hline
 \end{tabular}
\caption{The best-fit values for cosmological parameters and the $1\sigma$ limits from the extremal values of the N-dimensional distribution determined by the MCMC technique from the combined datasets including SN SDSS data with light curve fitting MLCS2k2 as well as HST and BBN. The current Hubble parameter $H_0$ is in units $km/(s\cdot Mpc)$, the age of the Universe $t_0$ is given in Giga years.}\label{tab1}
\end{table}

\section{Error forecasts for the Planck experiment}\label{futoc}

\begin{figure}
\centerline{\includegraphics[width=\textwidth]{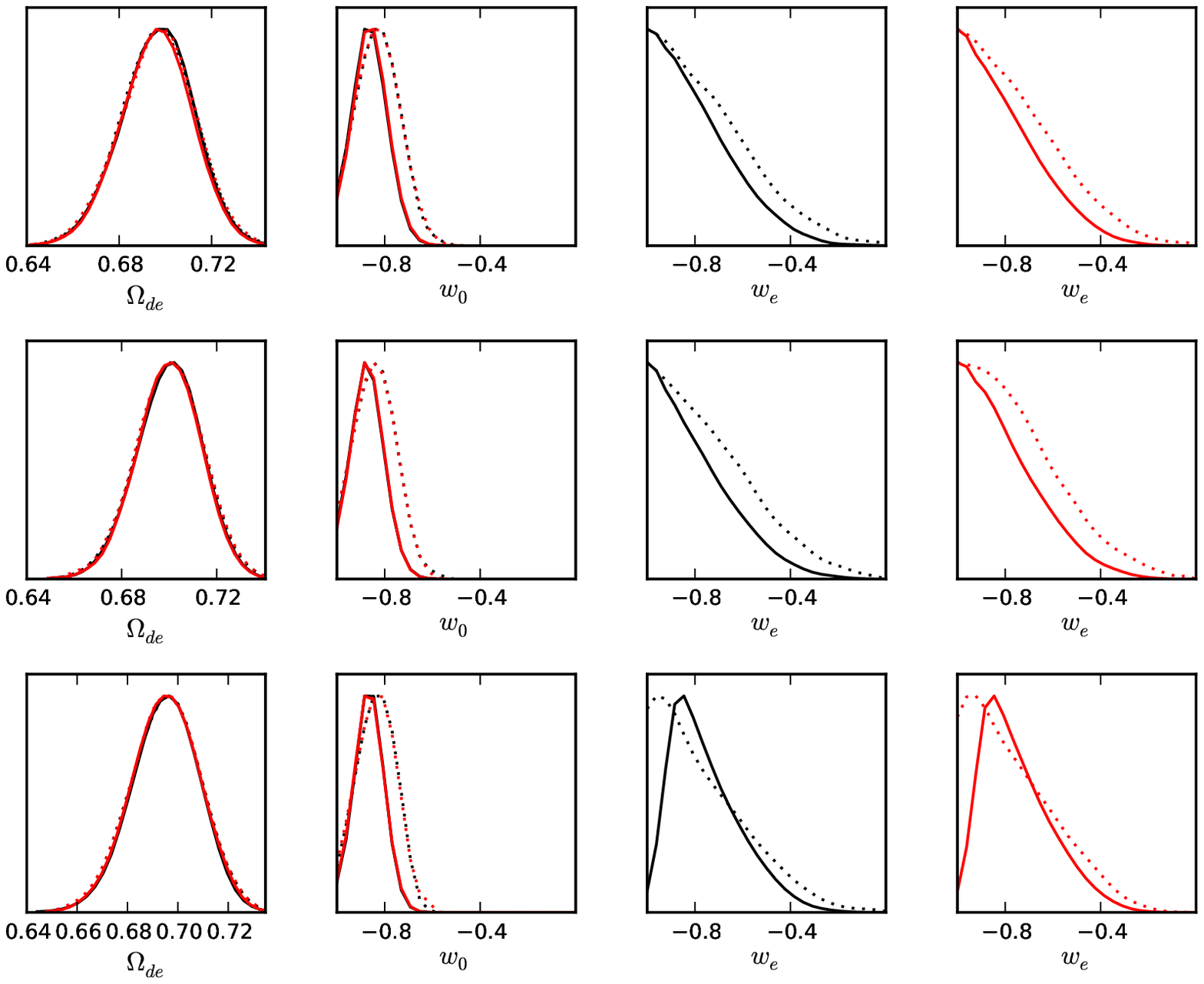}} 
\caption{One-dimensional marginalized posteriors (solid lines) and mean likelihoods (dotted lines) for the combined datasets Planck+HST+BBN+SN SDSS MLCS2k2, Planck+HST+BBN+SN SDSS MLCS2k2+BAO and Planck+HST+BBN+SN SDSS MLCS2k2+SDSS LRG7 (from top to bottom). From left to right: the functions for the dark energy density $\Omega_{de}$, the current value of dark energy EoS parameter $w_0$ for both CSF (black lines) and TSF (red), the early value of dark energy EoS parameter $w_e$ for CSF and the early value of dark energy EoS parameter $w_e$ for CSF.}
\label{fig12}
\end{figure}

\begin{figure}
\centerline{\includegraphics[width=0.65\textwidth]{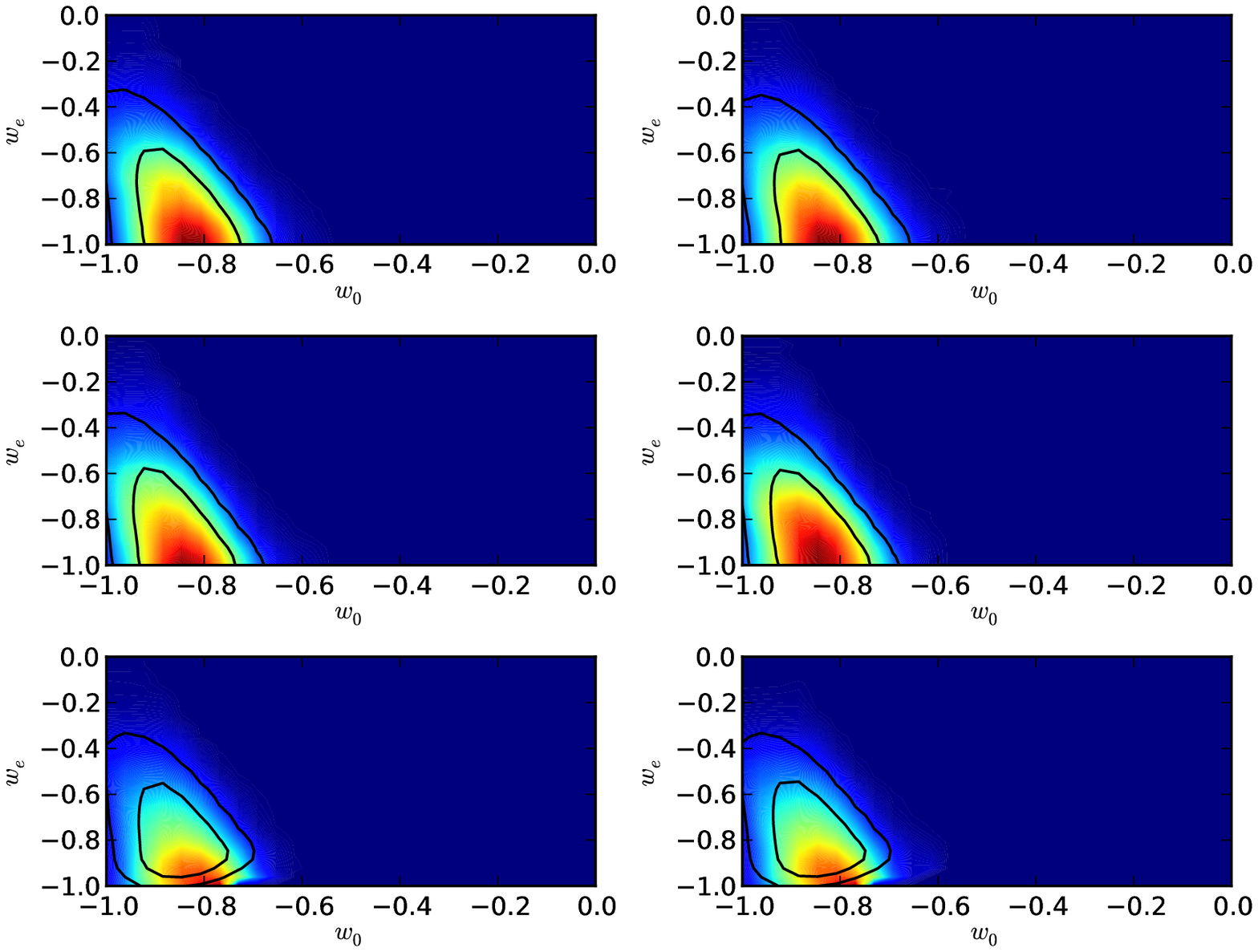}} 
\caption[]{Two-dimensional mean likelihood distributions in the plane $w_e-w_0$  for corresponding datasets and models from Fig.~\ref{fig12}. Solid lines show the $1\sigma$ and $2\sigma$ confidence contours.}
\label{fig13}
\end{figure}

In the previous section we have determined the observational constraints on cosmological parameters in models with scalar fields with barotropic EoS. Now we are going to discuss the precision, with which the expected Planck data on CMB anisotropies will allow us to estimate the parameters determining the barotropic EoS.

If the shape of the likelihood function cannot not be safely assumed to be Gaussian, the most reliable forecasting technique is the full MCMC analysis of mock data. To generate a Planck mock dataset we have used the publicly available code FuturCMB \cite{futurcmb_source}. The method of mock data generation and all necessary modifications of CosmoMC are thoroughly described in~\cite{futurcmb}. The fiducial $C_{\ell}$'s have been computed by CAMB for the set of best-fit parameters obtained for CSF from the dataset WMAP7+HST+BBN+SN SDSS MLCS2k2+SDSS LRG7 (Table \ref{tab1}, column 3). The seed has been chosen as 150. The generated mock dataset involves the CMB temperature fluctuations, polarization and the weak lensing deflection angle power spectra.

We assume that the Planck experiment has 3 channels, for which we choose in turn $\theta_{fwhm}$, $\sigma_T$ and $\sigma_E$ to be 9.5 arcmin, 6.8 $\mu K$ per pixel and 10.9 $\mu K$ per pixel; 7.1 arcmin, 6.0 $\mu K$ per pixel and 11.4 $\mu K$ per pixel; 5.0 arcmin, 13.1 $\mu K$ per pixel and 26.7 $\mu K$ per pixel correspondingly. The observed sky fraction is taken to be $f_{sky}=0.65$.

We have performed MCMC runs similar to those in the previous section and found that the Planck data alone as well as the combinations Planck+HST+BBN, Planck+HST+BBN\linebreak+BAO, Planck+HST+BBN+SDSS LRG7, Planck+HST+BBN+SN Union2 and Planck\linebreak+HST+BBN+SN SDSS SALT2 do not allow a reliable determination of the early value of EoS parameter. The likelihood for  $w_e$ is significantly non-Gaussian for any precision of the CMB data when the SN SDSS data with MLCS2k2 light curve fitting are not included.

In Fig.~\ref{fig12}-\ref{fig13} the one- and two-dimensional marginalized posteriors and mean likelihoods are presented for the datasets Planck+BBN+HST+SN SDSS MLCS2k2, Planck+BBN+HST\linebreak+SN SDSS MLCS2k2+BAO and Planck+BBN+HST+SN SDSS MLCS2k2+SDSS LRG7. Also in this case, the higher-redshift SN from the full SDSS compilation with the light curves fitted by MLCS2k2  reduce the non-Gaussianity of the likelihood for $w_e$. 

The substitution of the 7-year WMAP data on CMB  anisotropy by the Planck mock dataset reduces significantly the errors, so that the values of $w_e$ close to 0 appear to be far beyond the $2\sigma$ marginalized confidence contours. It is worth noting that from the two-dimensional marginalized distributions it follows that the $\Lambda$-term dark energy can be excluded at the $2\sigma$ confidence level for both Lagrangians and all datasets.

The best-fit values of the cosmological parameters for models with both fields and the corresponding $1\sigma$ limits from the extremal values of the N-dimensional distribution are presented in Table \ref{tab4}. The best-fit values are close to the values obtained from the corresponding datasets including WMAP7 data within $1\sigma$ limits, as expected. All best-fit models have $w_e<w_0$ as the fiducial model. Note that for the parameters $\Omega_bh^2$, $\Omega_{cdm}h^2$, $n_s$, $\log\left(10^{10}A_s\right)$, $z_{rei}$ and $t_0$ the presented in Table \ref{tab4} $1\sigma$ uncertainties are few times smaller than the corresponding uncertainties presented in Table \ref{tab1}. The uncertainties of determination of $w_e$ are significantly reduced by inclusion of the Planck mock data. The corresponding uncertainties of determination of the other dark energy parameters, $\Omega_{de}$ and $w_0$, as well as of the Hubble constant $H_0$ are also smaller than presented in Table \ref{tab1} ones.

\begin{figure}
\centerline{\includegraphics[width=0.9\textwidth]{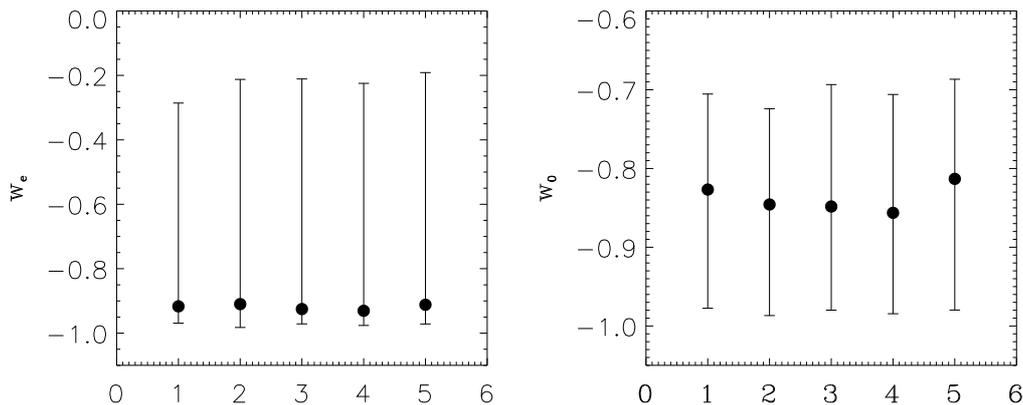}} 
\caption{The best-fit values of $w_e$ and $w_0$ along with $1\sigma$ limits from the extremal values of the N-dimensional distribution for 5 independent Planck mock datasets (with seeds 50, 100, 150, 200 and 250 from left to right in each panel).}
\label{fig14}
\end{figure}

Finally we want to check the reliability of the forecast. For this purpose we generate 4 additional independent Planck mock datasets with different seeds: 50, 100, 200 and 250. We have performed MCMC runs for these additional datasets combined with SDSS LRG7, SN SDSS MLCS2k2, HST and BBN. In Fig.~\ref{fig14} the best-fit values and $1\sigma$ limits from the extremal values of the N-dimensional distribution are shown for the parameters $w_0$ and $w_e$. We see that the best-fit values obtained from all datasets are within the $1\sigma$ confidence limits and the limits are generally consistent with each other. It can be stated with high confidence that the values $w_e>-0.1$ should be excluded by the combined datasets including forthcoming Planck data. This is consistent with our conclusion that the models with values of $w_e$ close to 0 could possibly be distinguishable from the corresponding models with $w_e$ close to $-1$ by the Planck data \cite{Novosyadlyj2010b,Novosyadlyj2010c}. Note that from Fig.~\ref{fig12}-\ref{fig13} and Table \ref{tab4} it can be deduced that CSF and TSF cannot be distinguished by CMB data from the next generation experiments since the differences between the models with both fields are smaller than the $1\sigma$ confidence limits. In the bottom row of Tables \ref{tab4} we see that the values of $-\log L$ differ for less than 1 for the best-fit parameters of models with CSF and TSF obtained from the same datasets. As it can be seen in Fig.~\ref{fig12}, the dark energy density $\Omega_{de}$ and its current value of EoS parameter $w_0$ are constrained well by all used combined datasets including Planck mock data and the curves for CSF and TSF overlap.

\begin{table}
 \begin{tabular}{|c|c|c|c|c|c|c|}
 \hline
 & & & & & &\\
 Parameters&CSF&CSF&CSF&
 TSF&TSF&TSF\\
 & & & & & &\\
 \cline{2-7}
 & & & & & &\\
 &PLANCK&PLANCK&PLANCK&PLANCK&PLANCK&PLANCK\\
  &SN SDSS&SN SDSS&SN SDSS&SN SDSS&SN SDSS&SN SDSS\\
 &MLCS2k2&MLCS2k2&MLCS2k2&MLCS2k2&MLCS2k2&MLCS2k2\\
 &&BAO&SDSS LRG7&&BAO&SDSS LRG7\\
 & & & & & &\\
 \hline
 & & & & & &\\
$\Omega_{de}$& 0.70$_{- 0.04}^{+ 0.04}$& 0.70$_{- 0.04}^{+ 0.04}$& 0.70$_{- 0.04}^{+ 0.03}$& 0.69$_{- 0.04}^{+ 0.04}$& 0.70$_{- 0.04}^{+ 0.03}$& 0.70$_{- 0.04}^{+ 0.03}$\\
 & & & & & &\\
$w_0$&-0.84$_{- 0.15}^{+ 0.19}$&-0.83$_{- 0.16}^{+ 0.17}$&-0.85$_{- 0.13}^{+ 0.15}$&-0.82$_{- 0.17}^{+ 0.18}$&-0.83$_{- 0.16}^{+ 0.15}$&-0.83$_{- 0.14}^{+ 0.14}$\\
 & & & & & &\\
$w_e$&-0.94$_{- 0.06}^{+ 0.72}$&-0.98$_{- 0.02}^{+ 0.86}$&-0.93$_{- 0.05}^{+ 0.71}$&-0.97$_{- 0.03}^{+ 0.76}$&-0.99$_{- 0.01}^{+ 0.77}$&-0.92$_{- 0.06}^{+ 0.71}$\\
 & & & & & &\\
$100\Omega_b h^2$& 2.30$_{- 0.04}^{+ 0.03}$& 2.29$_{- 0.03}^{+ 0.04}$& 2.29$_{- 0.03}^{+ 0.04}$& 2.29$_{- 0.03}^{+ 0.04}$& 2.29$_{- 0.03}^{+ 0.04}$& 2.29$_{- 0.03}^{+ 0.04}$\\
 & & & & & &\\
$10\Omega_{cdm} h^2$& 1.11$_{- 0.03}^{+ 0.03}$& 1.11$_{- 0.03}^{+ 0.03}$& 1.12$_{- 0.03}^{+ 0.03}$& 1.12$_{- 0.03}^{+ 0.02}$& 1.11$_{- 0.03}^{+ 0.03}$& 1.11$_{- 0.03}^{+ 0.03}$\\
 & & & & & &\\
$H_0$& 66.8$_{-  4.2}^{+  4.3}$& 66.7$_{-  3.4}^{+  4.1}$& 66.8$_{-  3.9}^{+  3.4}$& 66.3$_{-  3.9}^{+  4.7}$& 66.8$_{-  3.7}^{+  4.0}$& 66.5$_{-  3.7}^{+  3.8}$\\
 & & & & & &\\
$n_s$& 0.98$_{- 0.01}^{+ 0.01}$& 0.98$_{- 0.01}^{+ 0.01}$& 0.98$_{- 0.01}^{+ 0.01}$& 0.98$_{- 0.01}^{+ 0.01}$& 0.98$_{- 0.01}^{+ 0.01}$& 0.98$_{- 0.01}^{+ 0.01}$\\
 & & & & & &\\
$\log(10^{10}A_s)$& 3.10$_{- 0.03}^{+ 0.03}$& 3.10$_{- 0.02}^{+ 0.03}$& 3.10$_{- 0.02}^{+ 0.03}$& 3.10$_{- 0.02}^{+ 0.03}$& 3.10$_{- 0.02}^{+ 0.03}$& 3.10$_{- 0.02}^{+ 0.03}$\\
 & & & & & &\\
$z_{rei}$& 10.9$_{-  1.1}^{+  1.1}$& 10.9$_{-  1.0}^{+  1.1}$& 10.9$_{-  1.0}^{+  1.1}$& 10.8$_{-  1.0}^{+  1.2}$& 10.8$_{-  1.0}^{+  1.1}$& 10.9$_{-  1.0}^{+  1.2}$\\
 & & & & & &\\
$t_0$& 13.7$_{-  0.1}^{+  0.2}$& 13.7$_{-  0.1}^{+  0.1}$& 13.7$_{-  0.1}^{+  0.2}$& 13.8$_{-  0.1}^{+  0.2}$& 13.7$_{-  0.1}^{+  0.1}$& 13.8$_{-  0.1}^{+  0.2}$\\
 & & & & & &\\
 \hline
& & & & & &\\
$-\log L$&3416.86&3417.87&3439.87&3416.91&3417.93&3439.85\\
& & & & & &\\
\hline
 \end{tabular}
\caption{The best-fit values of cosmological parameters and $1\sigma$ limits from the extremal values of the N-dimensional distribution determined by the MCMC technique from the combined datasets including SN SDSS data with light curve fitting MLCS2k2 and Planck mock data instead of WMAP7. All datasets include also HST and BBN.}\label{tab4}
\end{table}

\section{Conclusion}\label{concl}
We have constrained the parameters of cosmological models with classical and tachyonic scalar fields with barotropic equation of state as dark energy using combined datasets including the CMB power spectra from WMAP7, the Hubble constant measurements, the Big Bang nucleosynthesis prior, the baryon acoustic oscillations in the space distribution of galaxies from SDSS DR7, the power spectrum of luminous red galaxies from SDSS DR7 and the light curves of SN Ia from 2 different compilations: Union2 (SALT2 light curve fitting) and SDSS (SALT2 and MLCS2k2 light curve fittings). We have found that the parameter corresponding to the value of $w$ at early times, is essentially unconstrained by most of the currently available data due to the significant non-Gaussianity of the likelihood function for $w_e$. To determine the best-fit value and the $1\sigma$ confidence ranges of $w_e$ the combined datasets including SN data from the full SDSS compilation with MLCS2k2 fitting of light curves have to be used, since only these SN data reduce the non-Gaussianity sufficiently. In these cases the best-fit scalar fields have  increasing EoS parameters, their repulsion properties recede and the Universe turns into contraction. 

We have also forecasted the uncertainties of the estimation of cosmological parameters of the studied models from the combined datasets including the data from the Planck experiment.  We were especially interested in the precision, with which the Planck data will constrain the early EoS parameter value. We have found that the non-Gaussianity of the likelihood function with respect to $w_e$ is not reduced by the expected Planck data alone. For the combined datasets including Planck mock data and SN data from the full SDSS compilation with MLCS2k2 light curve fitting method it is concluded that the models with $w_e>-0.1$ should be excluded at the $2\sigma$ confidence level.

\begin{acknowledgments}
This work was supported by the project of Ministry of Education and Science of Ukraine (state registration number 0110U001385), research program ``Cosmomicrophysics'' of the National Academy of
Sciences of Ukraine (state registration number 0109U003207) and the SCOPES project No. IZ73Z0128040 of Swiss National Science Foundation. Authors also acknowledge the usage of CAMB, CosmoMC and FuturCMB packages and are thankful to Main Astronomical Observatory of NASU for the possibility to use the computer cluster for MCMC runs. We would like to thank the anonymous referees for useful comments and suggestions.
\end{acknowledgments}

\end{document}